\documentclass[aps,prfluids,preprint,amsmath,amssymb,onecolumn,superscriptaddress,showkeys]{revtex4-2}
\usepackage{graphicx}
\usepackage{bm}
\usepackage{natbib}
\usepackage{caption}
\usepackage{color}
\usepackage{subfig}
\usepackage{url}
\begin{document}
\title {Polymer drag reduction in surfactant contaminated turbulent bubbly channel flows}
\author{Daulet Izbassarov}
\affiliation{Department of Mechanical Engineering, Aalto University, FI-00076 Aalto, Finland.}
\author{Zaheer Ahmed} 
\affiliation{Department of Mechanical Engineering, Mehran University of Engineering and Technology, SZAB Campus, Khairpur Mir’s 66020, Sindh, Pakistan.}
\author{Pedro Costa} 
\affiliation{Faculty of Industrial Engineering, Mechanical Engineering and Computer Science,
University of Iceland, Hjardarhagi 2-6, 107 Reykjavik, Iceland.}
\author{Ville Vuorinen}
\affiliation{Department of Mechanical Engineering, Aalto University, FI-00076 Aalto, Finland.}
\author{Outi Tammisola}
\affiliation{Linn\'{e} Flow Centre and SeRC, KTH Mechanics, Stockholm, Sweden.}
\author{Metin Muradoglu}
\email{mmuradoglu@ku.edu.tr}
%\homepage{http://home.ku.edu.tr/~mmuradoglu/index.htm}
\affiliation{Department of Mechanical Engineering, Koc University, 34450 Istanbul, Turkey.}

\date{\today}
\begin{abstract}
Polymer additives are commonly utilized to manipulate bubbly flows in various applications. Here, we investigate the effects of clean and contaminated bubbles driven upwards (upflow) in Newtonian and viscoelastic turbulent channel flows. Interface-resolved direct numerical simulations are performed to examine sole and combined effects of soluble surfactant and viscoelasticity using an efficient 3D finite-difference/front-tracking method. The incompressible flow equations are solved fully coupled with the FENE-P viscoelastic model and the equations governing interfacial and bulk surfactant concentrations. The latter coupling is accomplished by a non-linear equation of state that relates the surface tension to the surfactant concentration. For Newtonian turbulent bubbly flows, the effects of Triton X-100 and 1-Pentanol surfactant are examined. It is observed that the sorption kinetics highly affect the dynamics of bubbly flow. A minute amount of Triton X-100 is found to be sufficient to prevent the formation of bubble clusters restoring the single-phase behavior while even two orders of magnitude more 1-Pentanol surfactant is not adequate to prevent the formation of layers. For viscoelastic turbulent flows, it is found that the viscoelasticity promotes formation of the bubble-wall layers and thus the polymer drag reduction is completely lost for the surfactant-free bubbly flows, while the addition of small amount of surfactant (Triton X-100) in this system restores the polymer drag reduction resulting in $25\%$ drag reduction for the $Wi=4$ case. 
\end{abstract}
\maketitle
\section{Introduction}
Multi-fluid/multi-phase turbulent flows are ubiquitous in many natural processes and engineering applications, for instance, rain droplet and mist in environmental flows, food processing, bubble column reactors and water vapor bubbles in nuclear reactor cooling systems. The recent advancements in high-performance computing, and development of efficient numerical methods have made it possible to perform direct numerical simulations (DNS) of multiphase turbulent flows. DNS of multiphase flows are far more challenging than DNS of single-phase turbulent flows. In addition to the inherent difficulties arising from a wide range of length and time scales in turbulent flows, the existence of evolving interfaces, the effects of surface tension and discontinuous variation of the material properties across the phase boundaries, make these flows extremely complicated. The intricate multiphysics effects such as surfactant and viscoelasticity add further complexity and pose a computationally challenging problem. Addition of surfactants and polymers introduces elasticity (i.e., memory effects) to the interface and the bulk fluid, respectively. It is known that addition of small amount of surfactant can alter the behavior of bubbly flows completely~\cite{takagi2011surfactant}. The long-chain surfactant molecules do not only modify the interfacial properties (e.g., reduce the surface tension) but they also make the host fluid viscoelastic~\cite{tamano2014turbulent}. Moreover, long-chain polymers are often added to turbulent liquid flows for the purpose of drag reduction (DR) known as Toms effect~\cite{Toms-1949,White}. The drag-reducing polymer molecules also act as a surfactant in multiphase flows. It is therefore fundamentally important to understand the interactions and combined effects of the surfactant and viscoelasticity on turbulent multiphase flows from both fundamental and application points of view.

Addition of polymers to single-phase turbulent flows leads to a significant drag reduction. Since its discovery by Toms~\cite{Toms-1949}, drag reduction by polymer additives has been extensively studied for single-phase flows. Such a phenomenon is very sophisticated and its full understanding remains elusive~\cite{White}. Earlier explanations for the drag reduction can be classified into two schools of thought, one focusing on viscous effects~\cite{l2004drag,lumley1969drag,ryskin1987turbulent}, and the other on elastic effects~\cite{joseph2013fluid,tabor1986cascade}. Even though both theories have found experimental support, they have not achieved unequivocal conclusions. The drag reduction can be further categorized into two distinct regimes known as low drag reduction (LDR) and high drag reduction (HDR)~\cite{warholic1999influence}. \citet{zhu2018distinct} have shown that the polymer-induced drag reduction follows different mechanisms for these different regimes. For the LDR regime (DR$\leq40\%$), near-wall vortices are suppressed and streamwise velocity fluctuations are enhanced~\cite{dubief2004coherent,dubief2013mechanism}. However, the statistical trends are similar to Newtonian flow, i.e., the mean velocity profile follows a similar log-law slope with qualitatively similar velocity fluctuation profiles. On the other hand, for HDR ($40\%<$DR$\leq60\%$), streamwise velocity fluctuations almost vanish, Reynolds shear stress becomes significantly smaller and the slope of the log-layer mean velocity is dramatically modified. Furthermore, there is an upper limit on drag reduction referred as the maximum drag reduction asymptote (MDR) which depends on the Reynolds number ($Re$), i.e., the friction factor is given by $c_f^{-1/2}=19.0\log(Re \; c_f^{1/2})-32.4$~\cite{virk1970ultimate}. The recent study of \citet{choueiri2018exceeding} has shown that polymers can reduce the drag beyond the suggested asymptotic limit due to a re-laminarization of the flow. More detailed review on the topic can be found in \citet{xi2019turbulent}. Introducing a second, gaseous phase with surfactant contamination into this problem results in even more complex dynamics, as the two-fluid interface dynamics is tightly coupled with the elastic turbulence and the local surfactant concentration.

Direct numerical simulations (DNS) of bubbly flows in the absence of viscoelasticity and surfactants have been performed previously for both spherical bubbles~\cite{lu2013dynamics} and deformable bubbles~\cite{Lu08}. Simulations of pressure-driven turbulent bubbly channel flows~\cite{Lu06,Lu07,esmaeeli2005direct,dabiri2013transition,santarelli2016direct,bolotnov2011detached} have shown that the distribution of bubbles in the channel have significant effects on the overall flow dynamics. For instance, spherical bubbles tend to accumulate at the wall, forming bubble clusters that result in a significant reduction of the flow rate. \citet{lu2018direct,lu2019multifluid} have also performed DNS of turbulent bubbly flows undergoing topological changes. They examined the effects of various governing parameters on the evolution of flow statistics such as the average flow rate, wall shear, and interface area. The recent review by \citet{Elgobashi} encompasses the recent progress in the DNS of bubbly flows. Clearly, the improvement of conventional closure models for turbulent multiphase flows to be used in Reynolds-averaged, Favre-averaged or filtered (large-eddy) simulations of the Navier-Stokes equations~\cite{bois2017direct,magolan2017multiphase,feng2019bubble} demands high-quality interface-resolved DNS data.

Numerous industrial applications involve bubbly flows, for example, bubble column reactors and light water reactors. In these applications, contaminants such as surfactants are naturally present or sometimes deliberately added to manipulate the dynamics of bubbly flows. Indeed, the presence of surfactants has a drastic effect on the behavior of bubbly flows~\cite{takagi2011surfactant}. \citet{takagi2008effects} observed bubble clustering phenomena in an upward bubbly channel flow in the cases of clean and small enough surfactant concentrations. Conversely, the presence of a large amount of surfactants, or a small amount of a strong surfactant type (Triton-X100), prevented the formation of bubble clusters near the wall. \citet{Lu17} examined the effects of insoluble surfactant on turbulent bubbly flows by DNS and showed that the contaminated bubbles remain distributed across the whole channel. \citet{ahmed2020effects} have examined the effects of soluble surfactant on dynamics of a single bubble in a wide range of flow conditions. They found that the surfactant can dramatically change the bubble dynamics and reverse the direction of its lateral migration in a pressure-driven channel flow. The surfactant-induced Marangoni stresses act to move the bubble towards the channel center and the final position of the bubble is determined by the intricate interplay of the bubble deformability and interfacial surfactant concentration. \citet{soligo2019coalescence,soligo2020effect} developed a modified phase field method (PFM) for simulations of turbulent flows with large and deformable surfactant-laden droplets. They used this method to examine breakage/coalescence rates and size distribution of surfactant-laden droplets in turbulent flow. They showed that addition of surfactant can hinder bubble/droplet coalescence and thus significantly influences the final bubble/droplet distribution.

The interaction of a dispersed phase with turbulent flows of viscoelastic fluids is the subject of active research due to its relevance to numerous environmental and engineering applications. Recently several DNS works have considered viscoelastic turbulent channel flow laden with neutrally buoyant spherical rigid particles. \citet{esteghamatian2019dilute} studied dilute suspensions. They observed that, at low Weissenberg numbers ($\mathrm{Wi}$), the drag reducing effect of viscoelasticity is significantly attenuated. On the other hand, at high Weissenberg numbers the particles did not show any effect on the drag. Later, \citet{esteghamatian_zaki_2020} found that at moderate particle concentrations, the drag showed non-monotonic trend with $\mathrm{Wi}$. \citet{rosti2020increase} found that the drag reducing effect of polymer additives is completely lost for semi-dense suspensions. They also observed that the drag increases more for suspensions in viscoelastic fluids than for suspensions in Newtonian fluids. 

On the other hand, DNS studies on turbulent bubbly flow with combined effects of viscoelasticity and surfactants are scarce. To the best of our knowledge, \citet{ahmed2020turbulent} was the first study analyzing the interplay of surfactant and viscoelasticity on the behavior of vertical turbulent bubbly flows at $\mathrm{Re}_{\tau}=127$. It was found that the viscoelasticity promotes the formation of bubbly wall-layers and consequently the flow rate is reduced (i.e., drag increases). The formation of bubble wall-layers was related to the interplay of inertial, elastic and Marangoni forces. However, their work was focused on the numerical method and the parallelization algorithm, so they presented only a few results for the combined effects of surfactant and viscoelasticity at a low Reynolds number in a short channel with a small number of bubbles. In addition, they did not consider the effects of sorption kinetics of surfactant. Therefore, a further investigation is needed to reveal the combined effects of viscoelasticity and surfactants on turbulent bubbly flows. In the present study, we address this question and systematically examine the sole and combined effects of soluble surfactant and viscoelasticity on the dynamics of turbulent bubbly flows. For this purpose, extensive simulations are performed at $\mathrm{Re}_{\tau}=180$, with a larger number of bubbles ($N_b=133$). In addition, the realistic physical properties of Triton-X100 and 1-Pentanol surfactants are used to investigate the effects of their kinetic properties on the dynamics of turbulent bubbly flows. The main objectives of the present study are 
(i) to study effects of contamination on Newtonian turbulent bubbly flow, 
(ii) to examine the effects of clean bubbles on polymer drag reduction of viscoelastic turbulent base flow, 
(iii) to evaluate the effects of contaminated bubbly flow on polymer drag reduction of viscoelastic turbulent base flow,
and (iv) to analyze the velocity fluctuations and stress balance of bubbly flows in the absence and presence of viscoelastcity and surfactant. 

This paper is organized as follows. In the next section, we briefly describe the flow equations, the numerical method and the computational setup. Subsequently, in Section~\ref{sec:results}, we report and analyse the results of our simulations, starting with a Newtonian suspending liquid laden with contaminated bubbles, followed by the cases where the viscoelasticity effects are also considered. Finally, the conclusions are drawn in Section~\ref{sec:conclusions}.

\section{Numerical approach}\label{sec:numerics}

\subsection{Problem statement and computational setup}
\begin{figure}
\centering
\includegraphics[width=0.5\textwidth]{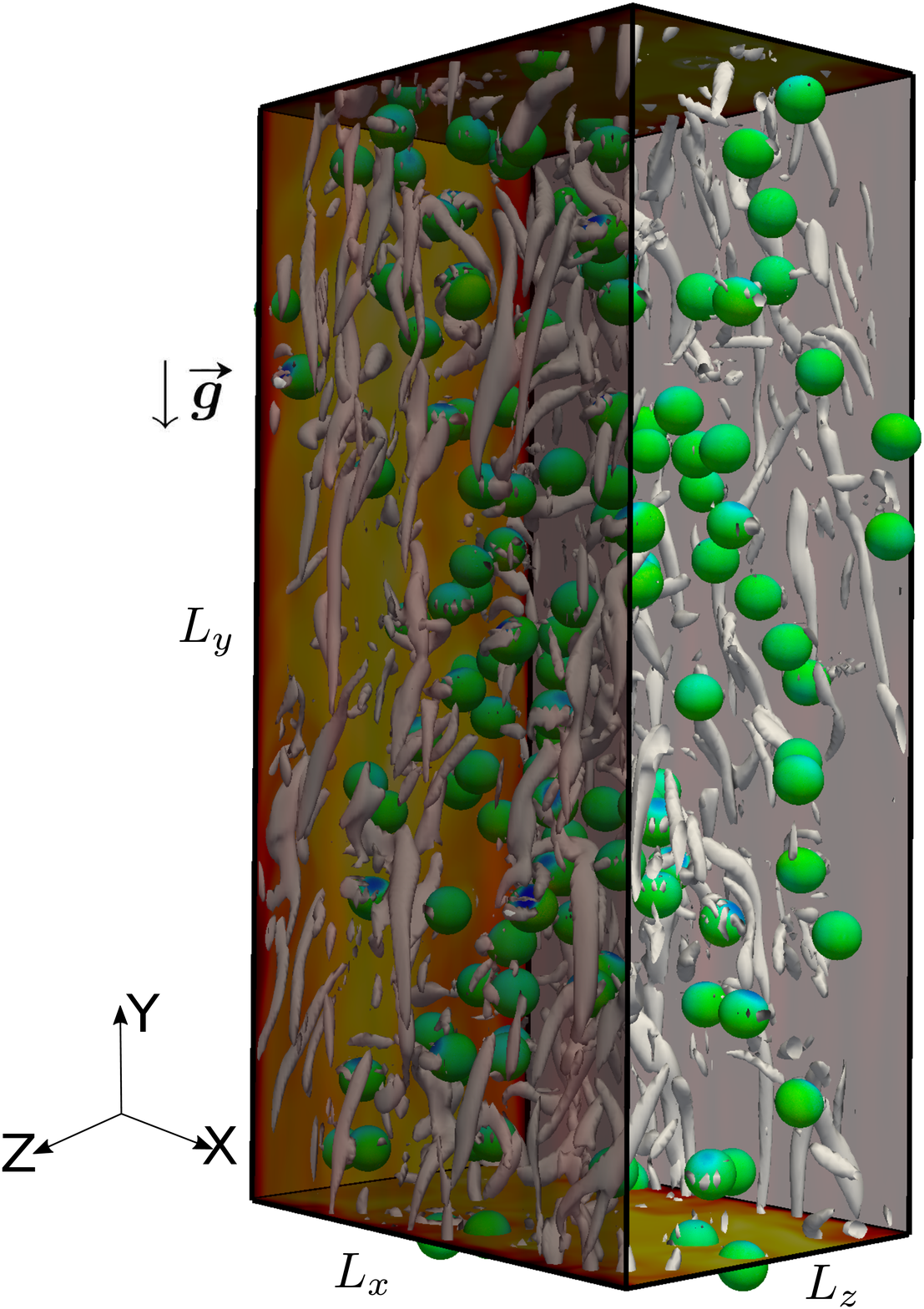}
\caption{Schematic representation of the computational setup considered in the present work. The flow is forced in the opposite direction to the gravity resulting in an upflow. The vortical structures are visualized using the $Q$-criterion, i.e., the iso-surfaces of the normalized second invariant of the velocity-gradient tensor, $Q^*/(V^*_b/(h^*))^2=2.5$, are plotted. The contours on the plane represent streamwise velocity with the scale ranging from $V^*_b=0$ m/s (red) to $V^*_b=20$ m/s (yellow). The contours on the bubble surface represent interfacial surfactant concentration with the scale ranging from 0 (blue) to 0.23 (green).}
\label{setup}
\end{figure}

 The computational domain illustrated in Fig.~\ref{setup} corresponds to a vertical plane channel with streamwise, spanwise and wall-normal directions along the $y$, $x$ and $z$ coordinates, respectively. The flow is periodic in the streamwise and spanwise directions, while no-slip and no-penetration boundary conditions are applied on the walls. The size of the domain $(L_y \times L_x \times L_z)$ is $6h \times 3h \times 2h$ and is resolved with $N_y \times N_x \times N_z =$ $576 \times 288 \times 240$ grid points in streamwise, spanwise and wall-normal directions respectively; $h=1$ is the half channel width. The grid points are uniformly spaced in the homogeneous directions and clustered close to the walls in the wall-normal direction using the following hyperbolic tangent mapping function
\begin{eqnarray}
z(k)= 1+ \frac{\tanh\big(\gamma(\frac{2k}{N_z}-1)\big)}{\tanh(\gamma)},
\end{eqnarray}
where the stretching parameter $\gamma=1.43$ is used. The resulting inner-scaled resolution is $\Delta x^+ =\Delta y^+ = 1.875$ and $ 0.498 \leq \Delta z^+ \leq 2.4$.

As the initial conditions, spherical bubbles are randomly placed in a single-phase, fully-developed turbulent flow with a friction Reynolds number of $\mathrm{Re}_\tau = v^*_{\tau} h^*/\nu^*_o = 180$. Here, the friction velocity is defined as $v^*_\tau = \sqrt{\tau^*_{wall} /\rho^*_o}$, where $\rho^*_o$ and $\nu^*_o$ are the density and kinematic viscosity of the suspending liquid, respectively. Note that, hereafter, unless otherwise stated, the superscript $^*$ is used to denote dimensional quantities, and variables without the superscript are non-dimensional (e.g., $p^*$ and $p$ represent the dimensional and non-dimensional pressure, respectively). The flow is driven by an applied constant pressure gradient ${\mathrm{d}p_0^*}/{\mathrm{d}y^*}$. At a statistically steady state, the average wall shear stress $\tau^{*}_{wall}$ is related to the pressure gradient and the weight of the bubble/liquid mixture through the streamwise momentum balance: 
\begin{equation}
\tau^{*}_{wall} = - \left(\frac{\mathrm{d}p_0^*}{\mathrm{d}y^*}+\rho^*_{av}g^*\right)h^*=-\mathcal{B^*} h^*,
\label{tau}
\end{equation}
where $\rho^*_{av}$ is the average density of the system (i.e., the total mass divided by the domain volume) and $h^*$ is the channel half-width. Note that the effective pressure gradient $\mathcal{B^*}$ determines bulk flow direction, i.e., upflow ($\mathcal{B^*}<0$) and downflow ($\mathcal{B^*}>0$). The present case pertains to upflow.

A swarm of mono-dispersed bubbles ($N_b=133$) is considered in this study, corresponding to a void fraction of $3\%$. The bubbles are initialized with a spherical shape with a diameter $d_b^*=0.25$ (in viscous units, $45.5 \nu/u_\tau$) and E\"{o}tv\"{o}s Number $\mathrm{Eo}= \rho^*_o g^*d^{*2}_{b}/\sigma^*_s = 0.45$. The Morton number ($M= g^*{\mu^*_o}^{4}/\rho^*_o {\sigma^*_s}^{3}$) used here is $M=6.17\times 10^{-10}$, which is higher than the Morton number of $M = 2.52\times10^{-11}$ for an air bubble in water at $20^\circ$C, but could be matched by using an aqueous solution of sugar~\cite{stewart1995bubble}. The physical parameters governing the flow are listed in Table~\ref{table:parametrs} where subscripts ``$i$" and ``$o$" denote the properties of the inner (dispersed) and the outer (continuous) fluids, respectively, and $\lambda^*$ is the polymer relaxation time. 

\begin{table}
\caption{Physical and computational parameters governing the flow.}
\label{table:parametrs}
\begin{tabular}{l l  l }
\hline
\hline
Density and viscosity ratio & \quad $\rho^*_i/\rho^*_o$, $\mu^*_i/\mu^*_o$    & \quad 0.02, 0.02   \\
Effective pressure gradient & \quad $\mathcal{B^*}=\frac{dp_0^*}{dy^*}+\rho^*_{av}g^*$   & \quad -1 \\
Surface tension & \quad $\sigma_{s}^*$ & \quad 3.334    \\ 
Gravitational acceleration & \quad $g^*$& \quad -24    \\
Weissenberg Number & \quad $\mathrm{Wi}=\lambda^* V^*_b/h^*$ & \quad 0/4/8  \\
Solvent viscosity ratio & \quad $\beta=\mu^*_s/\mu^*_o$ & \quad 0.9 \\
Extensibility parameter & \quad $L$ & \quad 60 \\
\hline
\hline
\end{tabular}
\end{table} 

The sorption kinetic properties of Triton X-100 and 1-Pentanol surfactants, widely used in experimental studies~\cite{takagi2011surfactant}, are considered in the present work. The corresponding physical adsorption $(k^*_a)$ and desorption $(k^*_d)$ properties are taken from \citet{takagi2009surfactant}. For Triton X-100, $k^*_a=50\,\mathrm{m}^3\mathrm{mol}^{-1}\mathrm{s}^{-1}$ and $k^*_d=0.033\,\mathrm{s}^{-1}$, whereas for 1-Pentanol, $k^*_a= 5.08\,\mathrm{m}^3\mathrm{mol}^{-1}\mathrm{s}^{-1}$ and $k^*_d=110.24\,\mathrm{s}^{-1}$. It is clear that Triton X-100 is adsorped by the interface much faster and is desorped back into the bulk fluid much slowly than 1-Pentanol, which makes Triton X-100 a more effective surfacant as will be discussed in the results section. The non-dimensional numbers related to the surfactants physical properties are listed in Table~\ref{table:srfparametrs}; $C_\infty~(=\frac{C_\infty^*}{C^*_{ref}})$ and $\Gamma~(=\frac{\Gamma^*}{\Gamma^*_{max}})$ represent the non-dimensional initial bulk surfactant concentration and the non-dimensional interfacial surfactant concentration, respectively, where $C^*_{ref}$ denotes the reference bulk surfactant concentration, taken as the critical micelle concentration (CMC), and $\Gamma^*_{max}$ represents the maximum packing concentration. The $C^*_{ref}$ values for the Triton X-100 and 1-Pentanol used are 1 ppm and 100 ppm, respectively. We note that the actual values of the Peclet number are much larger than the values in Table~\ref{table:srfparametrs} for both surfactants but the Peclet numbers are kept small to avoid thin mass boundary layer at the interface that requires excessive grid resolution.

\begin{table}
\caption{Non-dimensional numbers for TritonX-100 $\&$ 1-Pentanol surfactant cases.}
\label{table:srfparametrs}
\begin{tabular}{llcc}
\hline
&  & TritonX-100 & \quad 1-Pentanol\\
\hline
Peclet number  & $Pe_c=Pe_s=h^*V^*_b/D^*_c$ &777  & \quad 777 \\
Biot number & $Bi=h^*k^*_d/V^*_b$  & $2\times 10^{-3}$ & \quad 7.0\\ 
Damkohler number & $Da=\Gamma^*_{max}/h^*C^*_{ref}$ &0.456 &\quad 0.0006\\ 
Langmuir number & $La=k^*_a C^*_{ref}/k^*_d$ &2.2  & \quad $0.052$\\
Elasticity number & $\beta_s=R^*T^*\Gamma^*_\infty/\sigma^*_s$    & 0.5& \quad 0.5\\ 
\hline
\end{tabular}
\end{table} 

The bubble viscous response time can be estimated from Stokes drag as $\tau^*_b={d^*_b}^2/36\nu^*_o= 0.3124$. To determine the Kolmogorov time scale of the turbulent flow $\tau^*_k=\sqrt{\nu^*_o/\epsilon^*}$, the turbulent energy dissipation rate $(\epsilon^*)$ can be estimated from the flow velocity field, or by $\epsilon^*=v_\tau^3/(\kappa z^*_b)$, where $\kappa=0.42$ is the von K\'arm\'an constant and $z^*_b$ is the bubble average distance to the wall~\cite{lu2013dynamics}. Assuming a characteristic average distance $z^*_b = 0.5$~\cite{lu2013dynamics}, the Stokes number is estimated to be $St=\tau^*_b/\tau^*_k = 9.14$.

\subsection{Numerical method}

The continuity and momentum equations for an incompressible viscoelastic fluid flow are discretized using a second-order finite-difference/front-tracking (FD/FT) method~\citep{Tryggvason}. In this method, a single set of governing equations is written for the entire computational domain, the variations of the material properties across the interfaces are taken into account and the effects of the interfacial surface tension are included in the momentum equations as a body force~\cite{Tryggvason,Muradoglu,Izbassarov15}. The governing equations are non-dimensionalized using a length scale $\mathcal{L^*}$, a velocity scale $\mathcal{U^*}$ and a time scale $\mathcal{T^*} =\frac{\mathcal{L^*}}{\mathcal{U^*}}$. The length and velocity scales are taken as $\mathcal{L^*}=h^*$ and $\mathcal{U^*}=V_b^*$ where $V_b^*$ is the average liquid velocity of the Newtonian single-phase flow, i.e., $V_b^*=15.56$. The density and viscosity are non-dimensionalized using the density $\rho^*_o$ and the total viscosity $\mu^*_{o}$ of the continuous phase while the surface tension is normalized by the surface tension of the surfactant-free gas-liquid interface $\sigma^*_s$. The non-dimensional momentum equation, accounting for the interphase coupling, is then given by
\begin{eqnarray}
\frac{\partial{\rho}{\mathbf{u}}}{\partial t}+\nabla\cdot({\rho}{\mathbf{u}}{\mathbf{u}})&=&-\nabla{p}-\frac{dp_0}{dy}{\mathbf{j}}+\frac{(\rho-\rho_{av}){\mathbf{g}}}{Fr^2} + \nabla\cdot{\boldsymbol\tau} 
+\frac{1}{Re}\nabla\cdot\mu_s(\nabla{{\mathbf{u}}}+\nabla{{\mathbf{u}}^T}) \nonumber \\
&+&\frac{1}{We}\int_A\big[\sigma(\Gamma)\kappa{\bf{n}}+\nabla_s\sigma(\Gamma)\big]\delta({\bf{x}}-{\bf{x_f}})dA,
\label{NS}
\end{eqnarray}
where $\rho$, ${\mathbf{u}}$, $p$, $\pmb{\tau}$, $\mu_s$ and $\sigma$ are the non-dimensional density, the velocity vector, the pressure, the polymer stress tensor, the solvent viscosity and the surface tension coefficient, respectively. The unit vector ${\mathbf{g}}$ points in the direction of the gravitational acceleration. In Eq.~(\ref{NS}), the last term on the right hand side represents the surface tension where $A$ is the surface area, $\kappa$ is twice the mean curvature, $\mathbf{n}$ is the unit normal vector and $\nabla_s$ is the gradient operator along the interface defined as $\nabla_s=\nabla-{\mathbf{n}}({\mathbf{n}}\cdot\nabla)$. 
\noindent The non-dimensional numbers in Eq.~(\ref{NS}) are the Reynolds number $(Re=\rho^*_o\mathcal{U^*}\mathcal{L^*}/\mu^*_o)$, the Froude number $(Fr=\mathcal{U^*}/\sqrt{g^*\mathcal{L^*}})$ with $g^*$ being the magnitude of the gravitational acceleration and the Weber number $(We= \rho^*_o\mathcal{U^*}{^2}\mathcal{L^*}/\sigma^*_s)$. We emphasize that the surface tension is a function of the interfacial surfactant concentration $\Gamma$ and $\nabla_s\sigma(\Gamma)$ represents the surfactant-induced Marangoni stress. The surface tension is related to the interfacial surfactant concentration via the modified Langmuir equation of state
\begin{eqnarray}
\sigma = {\rm Max}\left(\epsilon_{\sigma}, 1 + \beta_s \ln(1-\Gamma)\right),
\label{SurfEqState}
\end{eqnarray}
where $\epsilon_{\sigma}=0.05$ is used to limit excessive reduction in surface tension at high concentrations. For the surfactants used here, the value of $\epsilon_{\sigma}$ is typically higher than this value, e.g., $\epsilon_{\sigma}\approx 0.3$ for Triton X-100~\cite{Gobel-J-97-jcis} but it does not have any influence on the present results since $\Gamma$ remains much lower than the maximum packing concentration in all the simulations. The present soluble surfactant methodology is the same as that developed by~\citet{muradoglu2008front,Muradoglu} and have been extensively used in various previous studies~\cite{Olgac,Tasoglu,ahmed2020effects}. However, the method is briefly described here for completeness.

The evolution equation for the interfacial surfactant concentration has been derived by \citet{Stone}. In the front-tracking framework, it can be expressed in the non-dimensional form as 
\begin{align}
\frac{1}{A}\frac{D \Gamma A}{Dt}= \frac{1}{Pe_{s}}\nabla_{s}^2\Gamma+Bi\dot{S}_{\Gamma},
\label{Us}
\end{align}
where $A$ is the surface area of an element of the interface and $Pe_s=\frac{\mathcal{U^*}\mathcal{L^*}}{D^*_s}$ is the interfacial Peclet number with $D^*_s$ being the diffusion coefficient along the interface. The Biot number is defined as $Bi=\frac{k^*_d\mathcal{L^*}}{\mathcal{U^*}}$, where $k^*_d$ is the desorption coefficient. The non-dimensional source term $\dot{S}_\Gamma$ is given by
\begin{align}
\dot{S}_\Gamma=La C_s(1-\Gamma)-\Gamma,
\label{dotS}
\end{align}
where $C_s$ is the bulk surfactant concentration near the interface and $La$ is the Langmuir number defined as $La=\frac{k^*_a C^*_{ref}}{k^*_d}$ with $k^*_a$ being the adsorption coefficient. The bulk surfactant concentration is governed by an advection-diffusion equation of the form
\begin{align}
\frac{\partial C}{\partial t}+\nabla\cdot(C{\mathbf{u}})=\frac{1}{Pe_c}\nabla\cdot(D_{co}\nabla C),
\label{bulkconc}
\end{align}
where $Pe_c =\frac{\mathcal{U^*}\mathcal{L^*}}{D^*_{c}}$ is the Peclet number based on bulk surfactant diffusivity. The coefficient $D^*_{co}$ is related to the molecular diffusion coefficient $D^*_c$ and the phase indicator function $I$ as
\begin{align}
D^*_{co}=D^*_c I({\mathbf{x}},t).
\end{align}
The source term in Eq. \eqref{dotS} is related to the bulk concentration by
\begin{align}
\dot{S}_\Gamma=-\frac{1}{Pe_c Da}({\mathbf{n}}\cdot\nabla _\mathrm{s}C)_{\rm interface},
\label{sourceterm}
\end{align}
 where $Da=\frac{\Gamma^*_{max}}{\mathcal{L^*}C^*_{ref}}$ is the Damk\"{o}hler number. Following \citet{Muradoglu}, the boundary condition at the interface given by Eq.~\eqref{sourceterm} is first converted into a source term for the bulk surfactant evolution equation. In this approach it is assumed that all the mass transfer between the interface and bulk takes place in a thin adsorption layer adjacent to the interface. Thus, the total amount of mass adsorbed on the interface is distributed over the adsorption layer, and added to the bulk concentration evolution equation as a negative source term. Equation~\eqref{bulkconc} thus takes the following form:
\begin{align}
\frac{\partial C}{\partial t}+\nabla\cdot(C{\mathbf{u}})=\frac{1}{Pe_c}\nabla\cdot(D_{co} \nabla C)+\dot{S}_c,
\label{BulkSurf}
\end{align}
where $\dot{S}_c$ is the source term evaluated at the interface and distributed onto the adsorption layer in a conservative manner. The details of this treatment can be found in \cite{Muradoglu}.

The rheological properties of the suspending fluid are taken into account using the FENE-P model~\citep{Bird-etal-80-jnnfm} for the polymeric stress tensor $\pmb{\tau}$. The model equations can be written in the non-dimensional form as 
\begin{align}
\frac{\partial {\mathbf{B}}}{\partial t}+\nabla\cdot(\mathbf{u}{\mathbf{B}})-(\nabla \mathbf{u})^T\cdot \mathbf{B} -\mathbf{B}\cdot\nabla \mathbf{u}+\frac{1}{Wi}(F\mathbf{B}-\mathbf{I})=0, \quad
F=\frac{L^2}{L^2-\rm{trace}(\mathbf{B})}, \\
\pmb{\tau}=\frac{1}{ReWi}(1-\beta)(F\mathbf{B}-\mathbf{I}),
\end{align}
where $\mathbf{B}$, $\mathrm{Wi}$, $L$, and $\mathbf{I}$ are the conformation tensor, the Weissenberg number, the maximum polymer extensibility, and the identity tensor, respectively. In the present study, the viscoelasticity is characterized by the Weissenberg number defined as $\mathrm{Wi} = \frac{\lambda^*\mathcal{U^*}}{\mathcal{L^*}}$, where $\lambda^*$ is the polymer relaxation time. The viscoelastic constitutive equations are highly non-linear and notoriously difficult to solve at high Weissenberg numbers. In the present work, the log-conformation method is used~\cite{Fattal} to deal with the so-called high Weissenberg number problem. More details can be found in~\citep{Izbassarov15,Izbassarov18,ahmed2020turbulent}.

All the field equations are solved on a fixed Eulerian grid with a staggered arrangement where the velocity nodes are located at the cell faces while the material properties, the pressure, the bulk surfactant concentration and the extra stresses are all located at the cell centers. The interfacial surfactant concentration evolution equation is solved on a separate Lagrangian grid. The spatial derivatives are discretized with second-order central differences for the diffusive terms, while the convective terms are discretized using a QUICK scheme~\citep{Leonard} in the momentum equation, and a fifth-order WENO-Z~\citep{Borges} scheme in the viscoelastic and the bulk surfactant concentration equations. The equations are integrated in time with a second-order predictor-corrector method in which the first-order solution (Euler method) serves as a predictor that is then corrected by the trapezoidal rule~\cite{Tryggvason}. A pressure splitting technique \cite{Dong_JCP_2012} is employed to turn the variable-coefficient Poisson equation into a constant-coefficient pressure Poisson equation. The corresponding linear system is solved directly using the FFT-based direct Poisson solver~\cite{Costa} of the open-source DNS code \emph{CaNS}. More details of the present numerical method are given in~\citet{ahmed2020turbulent}.

\section{Results and Discussion}\label{sec:results}
Simulations are first performed to examine the effects of surfactant in the Newtonian turbulent bubbly flows using the physical properties of Triton X-100 and 1-Pentanol. Then we examine the interplay and combined effects of surfactant and viscoelasticity. We particularly investigate the effects of surfactant on polymer drag reduction in turbulent bubbly flows at fixed void fraction of $3\%$.

\subsection{Effects of surfactant}
Simulations of Newtonian bubbly flow have been performed for two types of surfactants, Triton X-100 and 1-Pentanol, with nondimensional concentration $C_\infty$ varying between 0 and 1. In addition, the effects of initial conditions of interfacial surfactant concentration $\Gamma$ are examined. More specifically, we consider initially clean bubbles with $\Gamma_{\textrm{initial}}=0$ and contaminated bubbles with $\Gamma_{\textrm{initial}}= \Gamma_{eq}$, where $\Gamma_{eq}$ is the interfacial surfactant concentration at equilibrium with the bulk conditions. More details of the different sets of simulations are given in Table~\ref{casenames}. 

\begin{table}[htb]
\caption{Specifications for surfactant-contaminated cases.}
\label{casenames}
\begin{tabular}{lclcccc}
\hline
\hline
Name &&   Surfactant && $\Gamma_{\rm initial}$ && $C_\infty$\\
\hline
\hline
Case~1  && Triton X-100  && $0$ && $0.05-1.0$\\
Case~2  && 1-Pentanol  && $0$ && $0.01$ and $1.0$\\
Case~3  && Triton X-100  && $\Gamma_{eq}$ && $0.1$, $0.25$ and $0.5$ \\
\hline
\hline
\end{tabular}
\end{table} 

\begin{figure}
\centering
\includegraphics[width=0.8\textwidth]{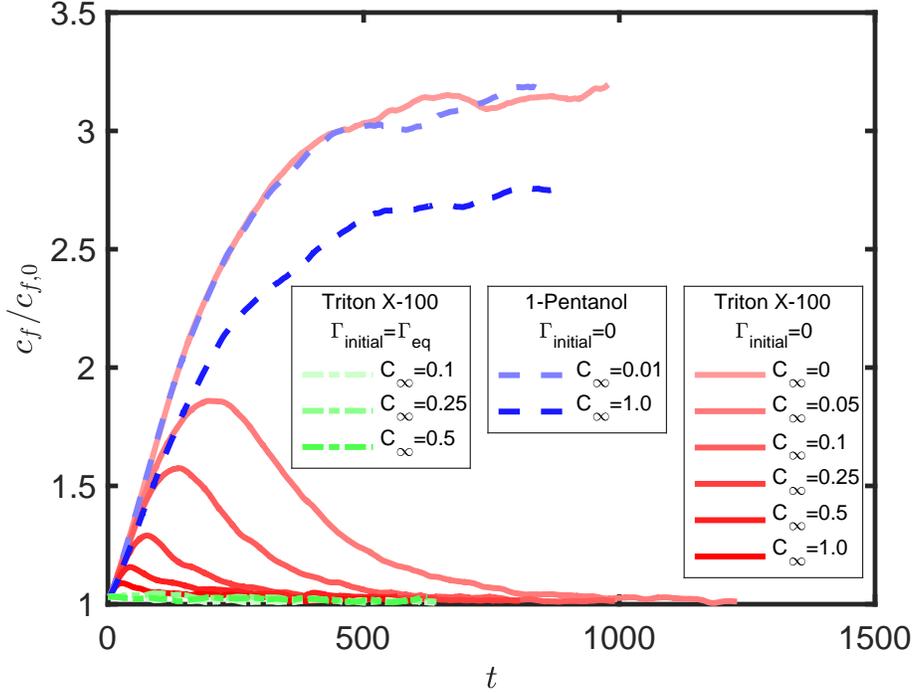}
\caption{Skin friction coefficient $(c_f)$ of the bubbly flow normalized by that of fully-developed single-phase flow $(c_{f,0})$. We note that Triton X-100 cases show low drag, while 1-Pentanol ones results in high values.}
\label{sfcN}
\end{figure} 

\begin{figure}
\centering
\subfloat[]{\includegraphics[width=0.48\textwidth]{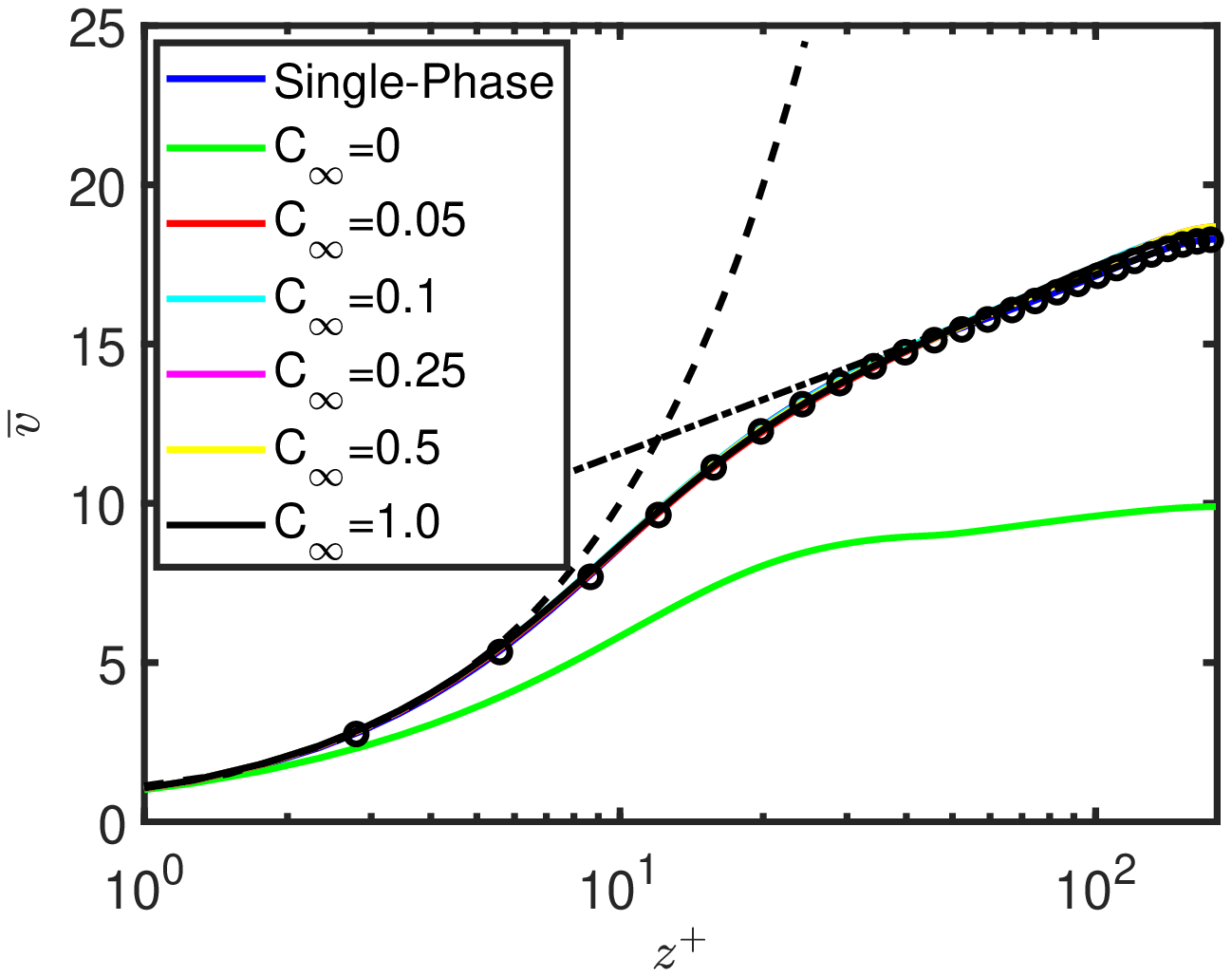}}
\subfloat[]{\includegraphics[width=0.48\textwidth]{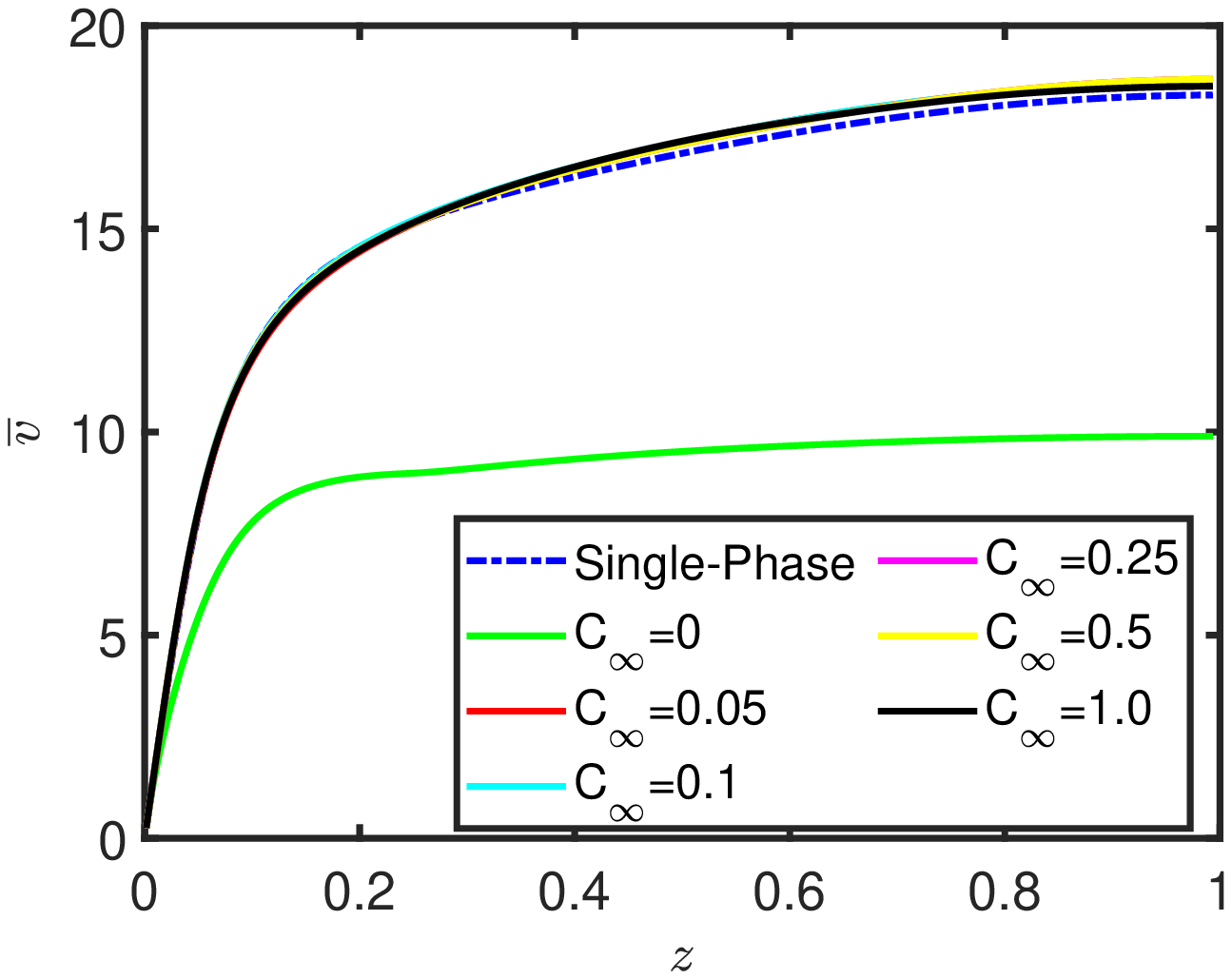}}\\
\subfloat[]{\includegraphics[width=0.48\textwidth]{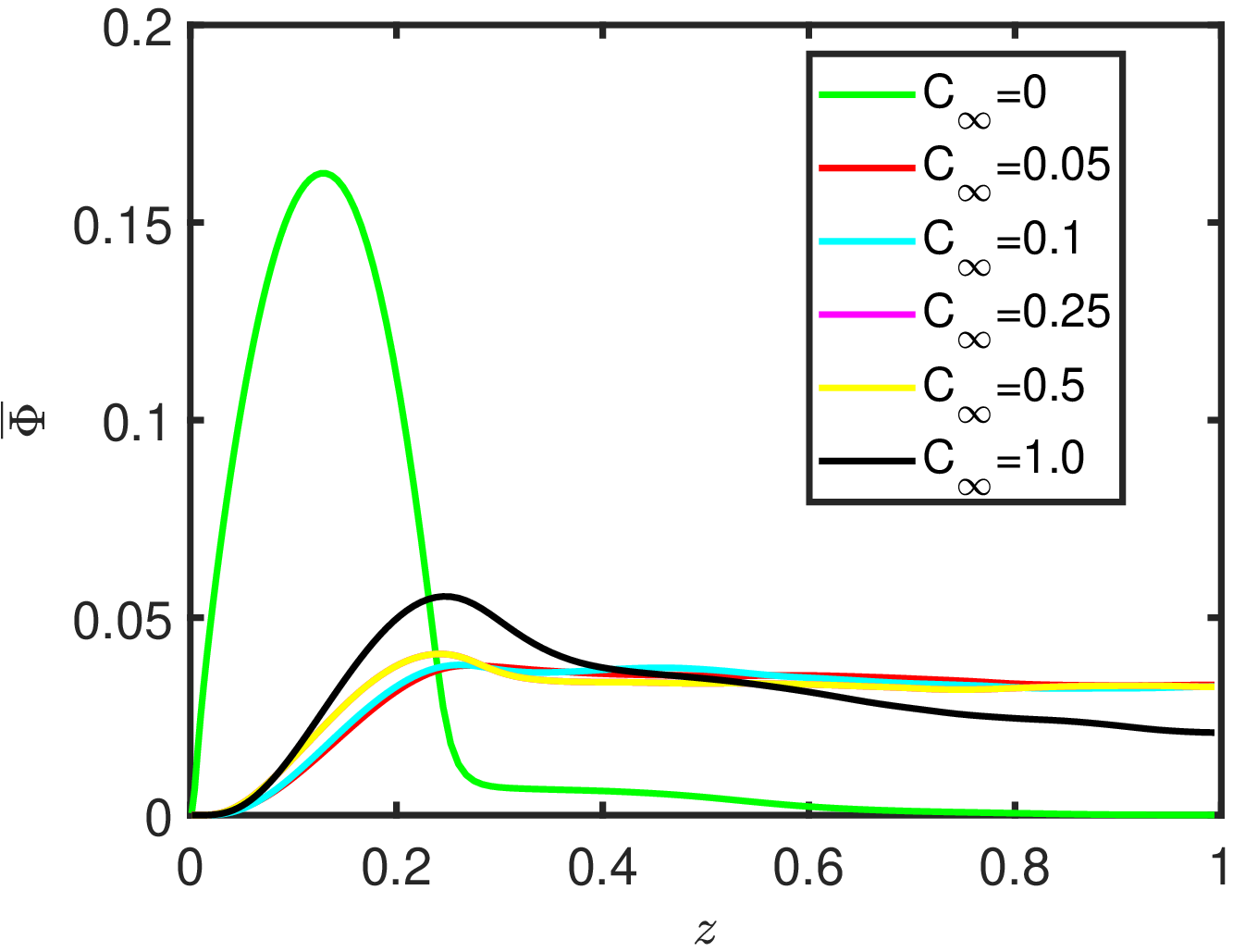}}
\caption{Surfactant Triton X-100: (a-b) Average liquid velocity scaled by $v_\tau$ vs wall-normal distance. The computed inner-scaled mean velocity profile (solid line) are compared with the law of the wall, i.e., $v^+ = z^+$ in the laminar sublayer (dashed line) and $v^+ = 2.5 \ln z^+ + 5.5$ in the logarithmic region (dash-dotted line). The solid lines represent the current results and the symbols are the DNS data of \citet{vreman2014comparison}. (c) Average void fraction at a statistically steady state.}
\label{velprof}
\end{figure}  

Figure~\ref{sfcN} shows the transient behavior of the skin friction coefficient $c_f=2\tau^*_{wall}/\rho^*_o{\overline{v}^*_o}^2$, with ${\overline{v}^*_o}$ being the average bulk liquid velocity, normalized by the value of the Newtonian single-phase flow, $c_{f,0} \approx 8.3\times10^{-3}$. For the clean case ($C_\infty=0$), $c_f$ increases monotonically, until it reaches a plateau of $c_f/c_{f,0} \approx 3.2$. Let us now consider the cases where initially clean bubbles are exposed to the bulk surfactant ($\Gamma_{\textrm{initial}}=0$). For Triton X-100, $c_f$ first increases reaching a peak, followed by a decrease approaching the value of the corresponding single-phase flow $c_f=c_{f,0}$ at the steady state. The value at the peak and the time to reach the steady state decrease with increasing $C_\infty$. For 1-Pentanol, however, the time-evolution of $c_f$ is similar to the clean case. Moreover, high surfactant concentrations are needed to cause a clear effect of 1-Pentanol. The curve for this surfactant at $C_\infty=0.01$ is almost identical to the clean case. When increasing the surfactant concentration further up to $C_\infty=1.0$, the behaviour for 1-pentanol remains qualitatively similar, with only difference that the upper limit of $c_f/c_{f,0}$ decreases to $\approx 2.6$. Finally, we see a large effect of the initial conditions for Triton X-100. The $\Gamma_{\textrm{initial}}= \Gamma_{eq}$ cases for Triton X-100 follow a nearly the corresponding single-phase case. The vast variation between different initial conditions could be attributed to the very low desorption rate of Triton X-100. Surfactant molecules at the leading part of bubble surface are transported to the trailing part of bubble surface. Thus, large surface concentration gradient generates strong Marangoni stresses which prevent the lateral migration of the bubbles toward the wall~\cite{ahmed2020effects,ahmed2020turbulent}.

\begin{figure}
\centering
\subfloat[]{\includegraphics[width=0.48\textwidth]{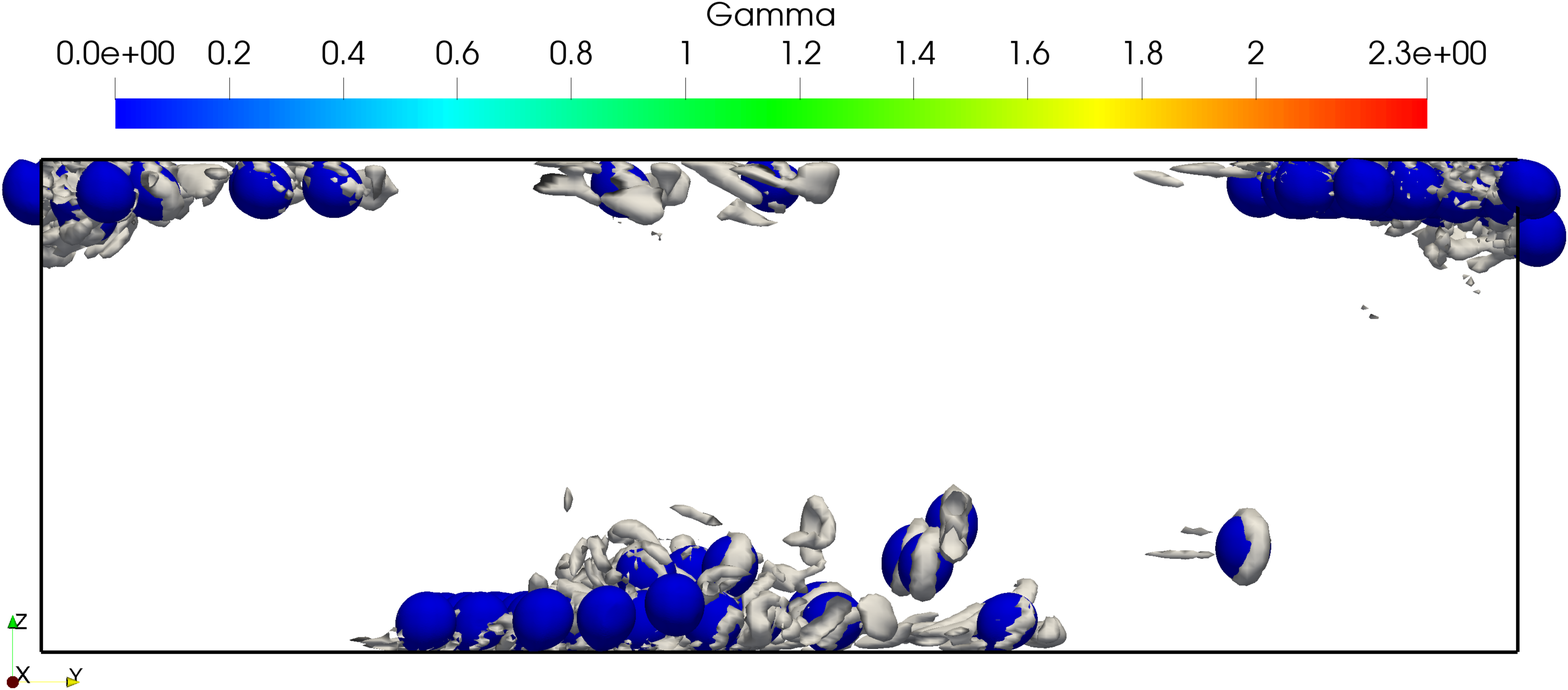}}
\subfloat[]{\includegraphics[width=0.48\textwidth]{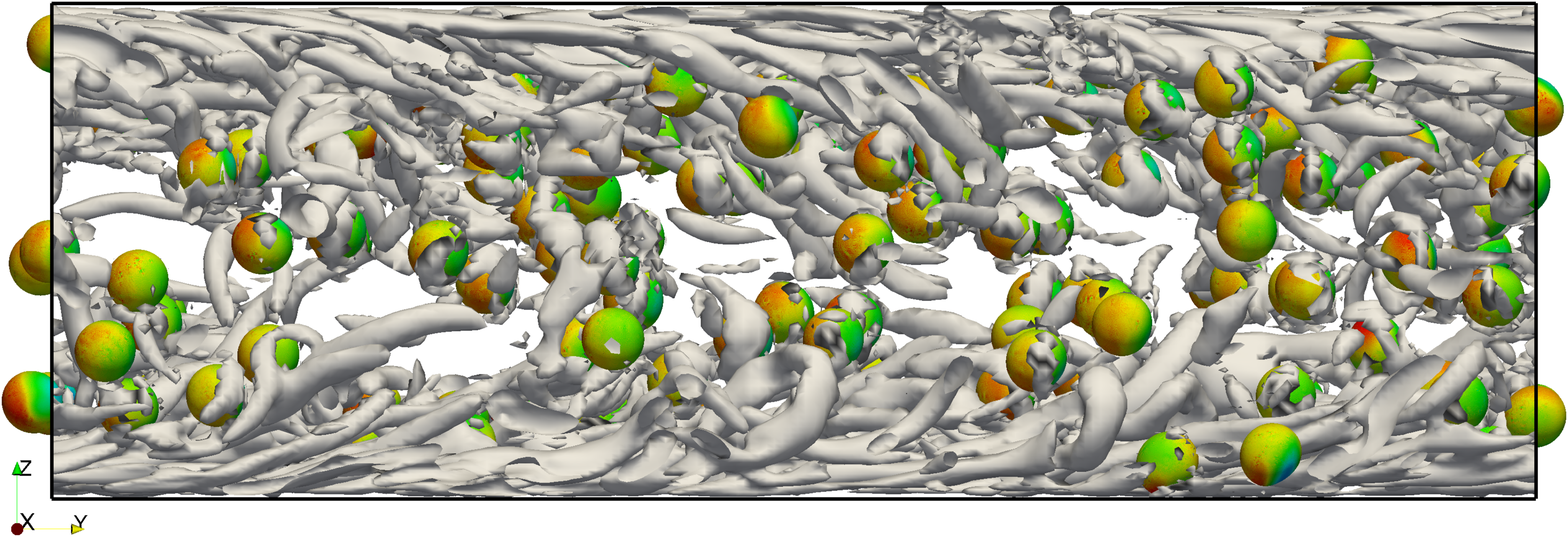}}\\
\subfloat[]{\includegraphics[width=0.48\textwidth]{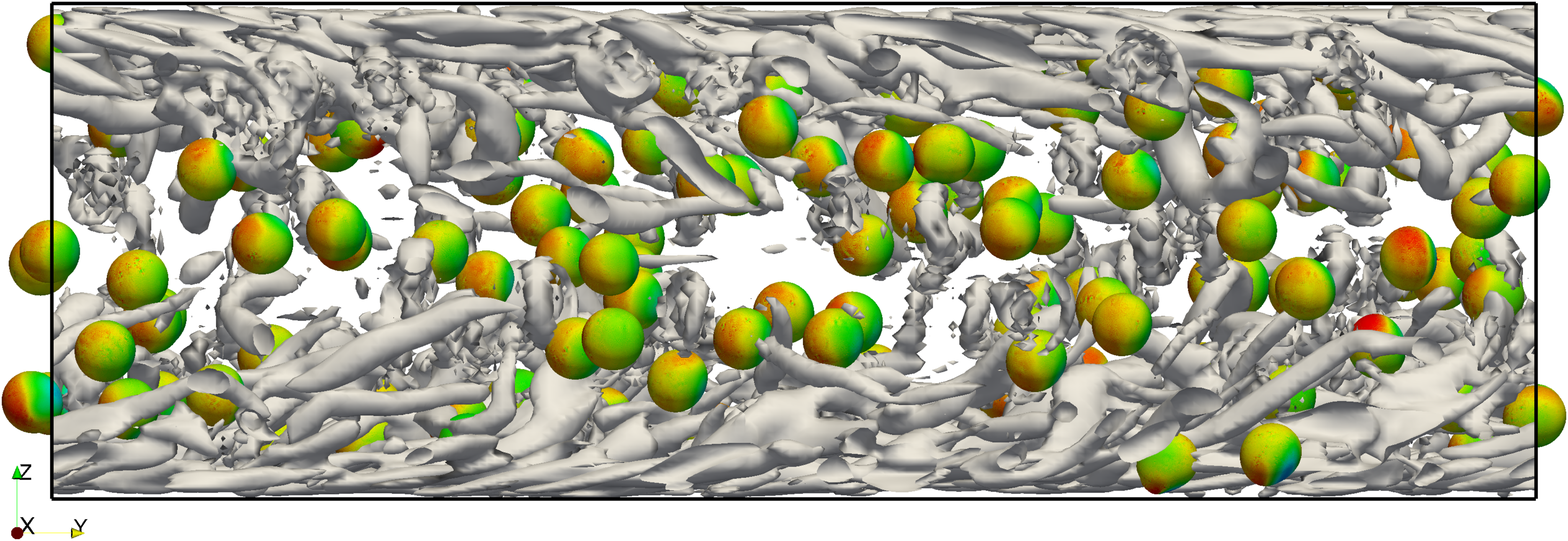}}
\subfloat[]{\includegraphics[width=0.48\textwidth]{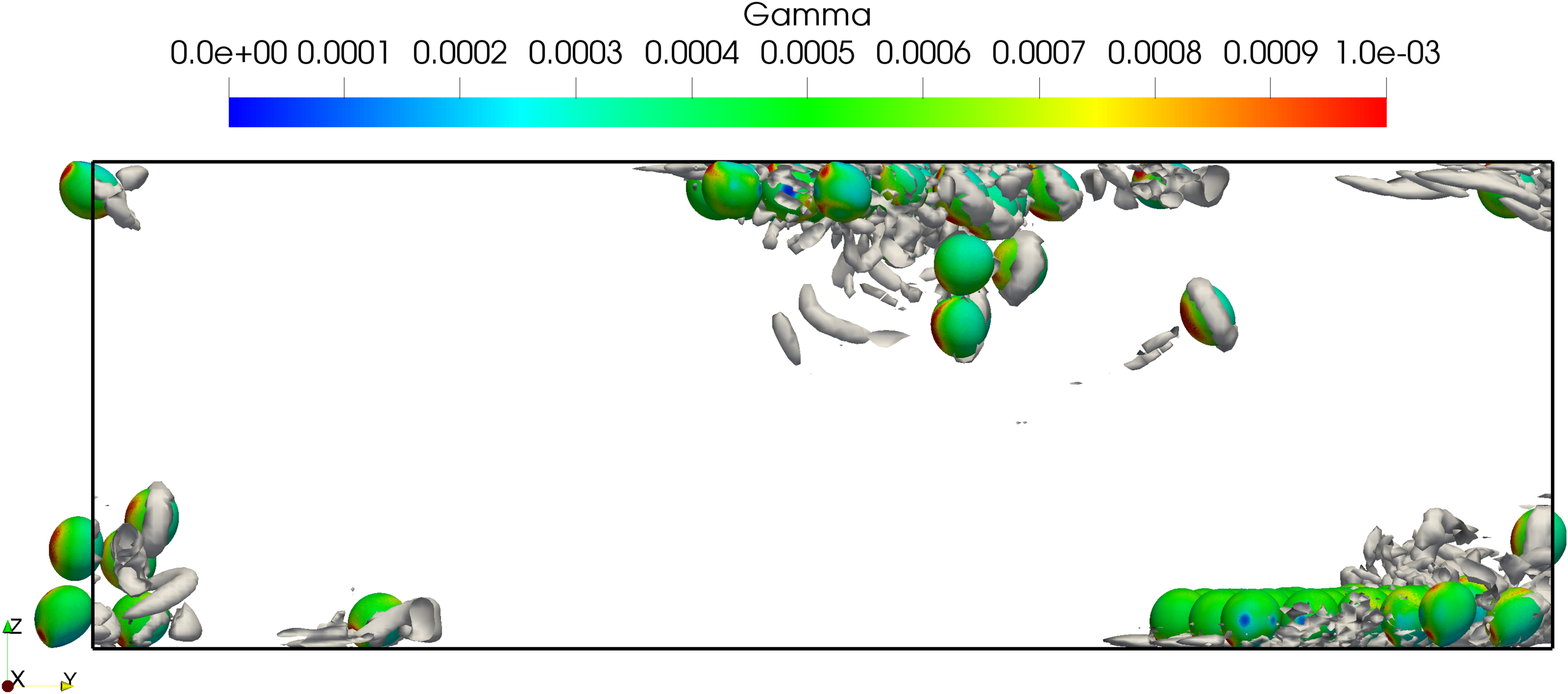}}
\caption{Newtonian Flow: Vortical structures at a statistically steady state for Triton X-100 (a) $C_{\infty}=0.0$ (b) $C_{\infty}=0.1$ (c) $C_{\infty}=1.0$ and for 1-Pentanol (d) $C_{\infty}=1.0$. The vortical structures are visualized using $Q$-criterion~\cite{jeong1995identification}, i.e., the iso-contours of the normalized second invariant of the velocity-gradient tensor, $Q^*/(V^*_b/(h^*))^2=0.8$, are plotted. The contours on the bubble surface represent interfacial surfactant concentration $(\Gamma)$.}
\label{QcritN}
\end{figure}

\begin{figure}
\centering
\subfloat[]{\includegraphics[width=0.5\textwidth]{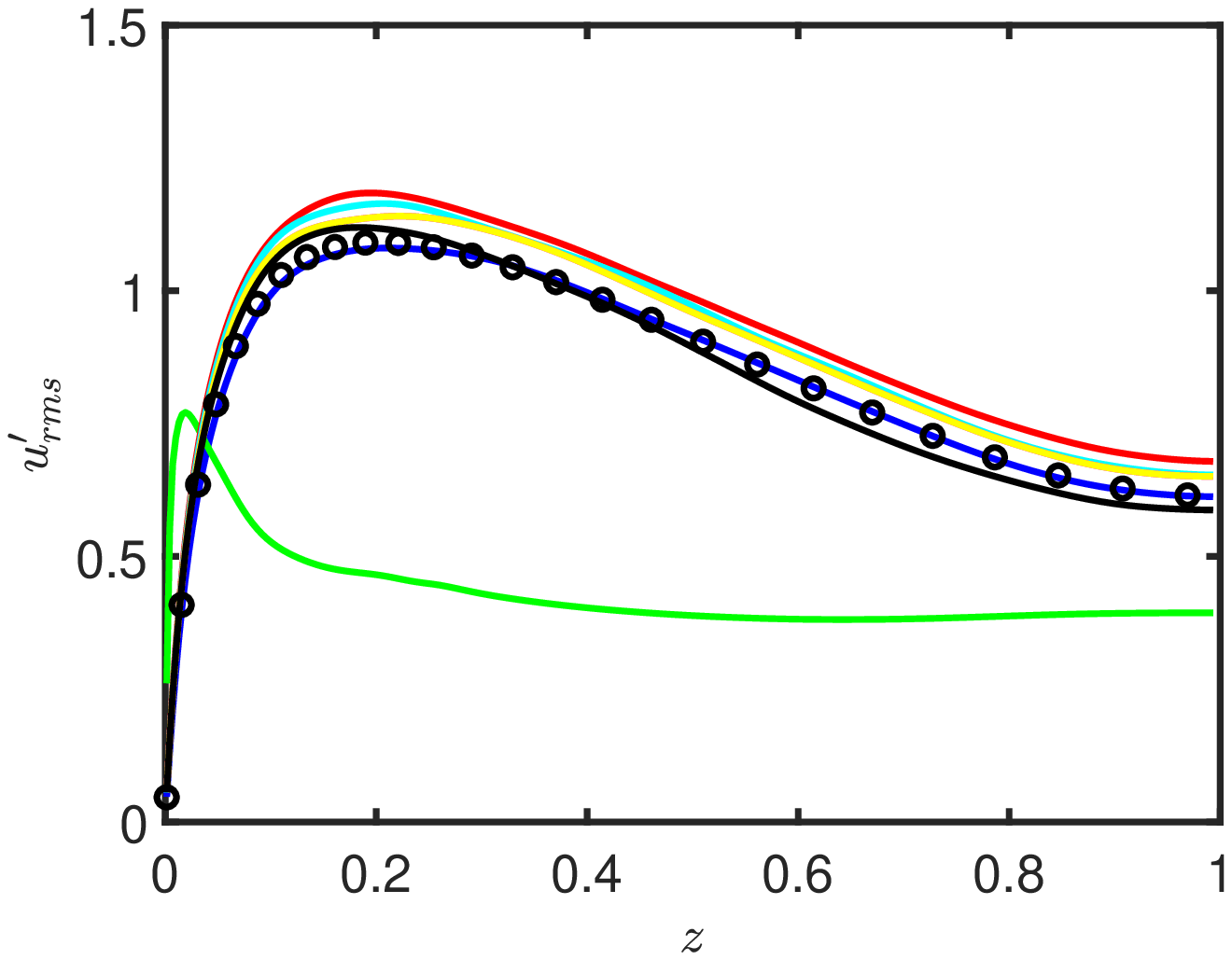}}
\subfloat[]{\includegraphics[width=0.5\textwidth]{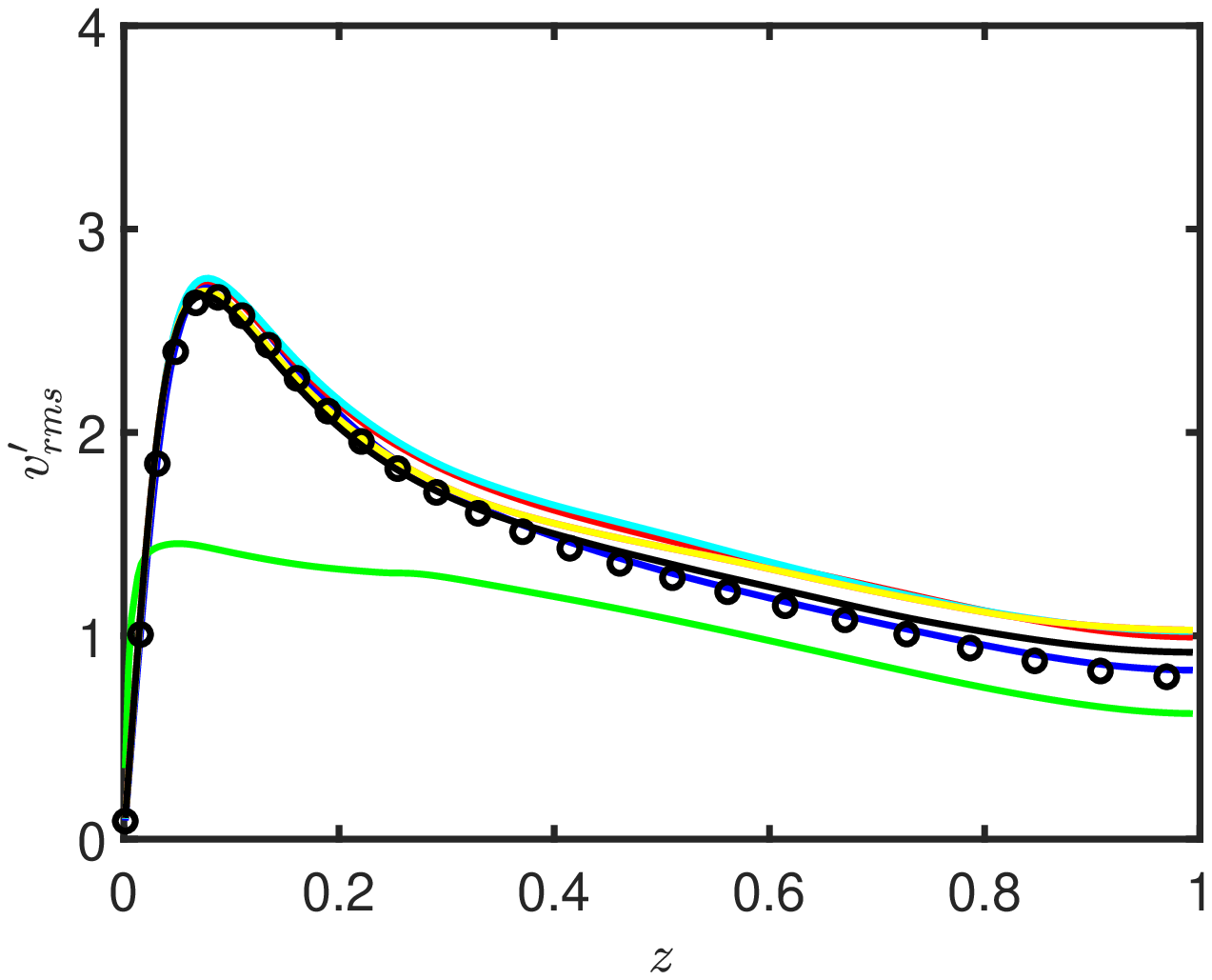}}\\
\subfloat[]{\includegraphics[width=0.5\textwidth]{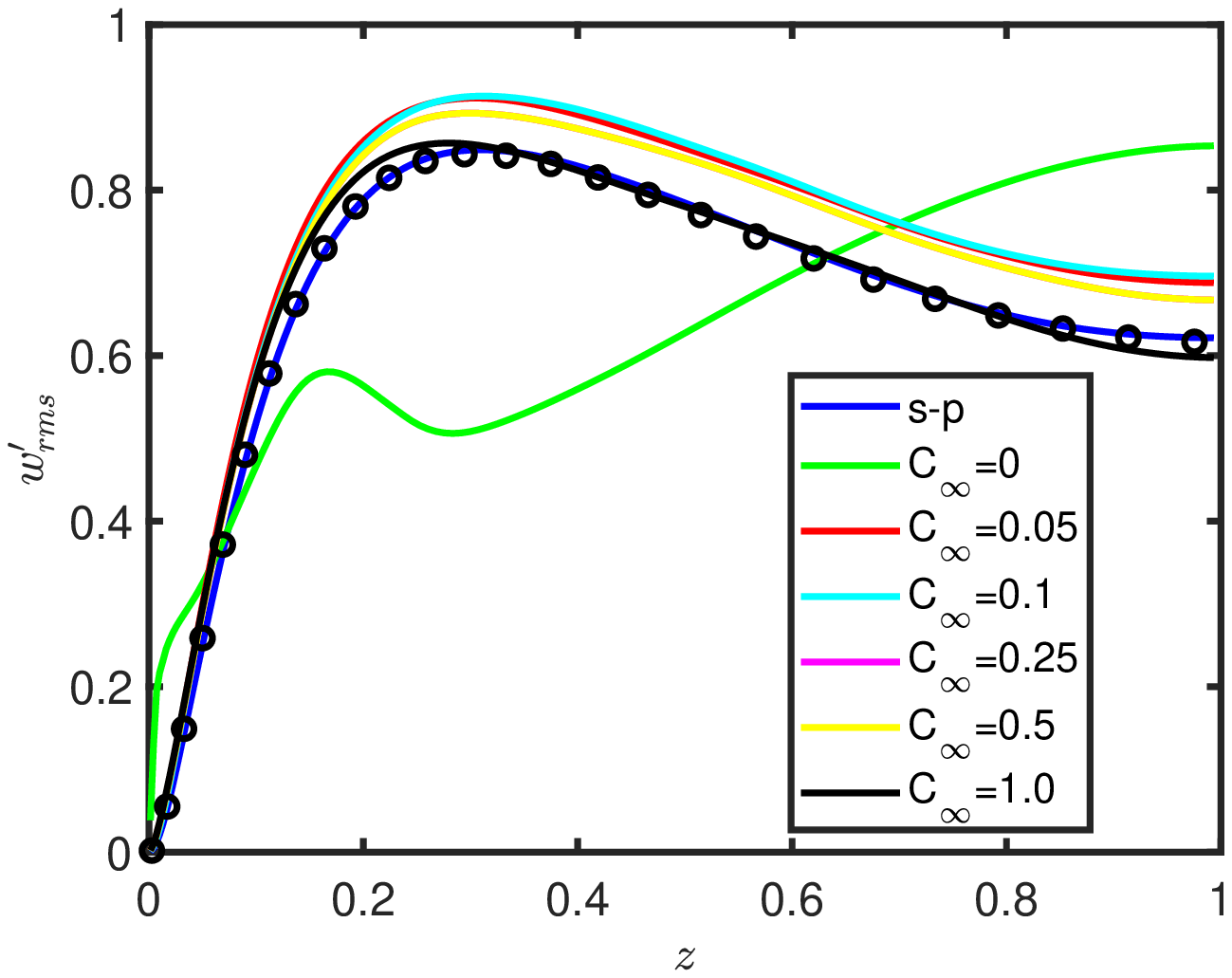}}
\subfloat[]{\includegraphics[width=0.5\textwidth]{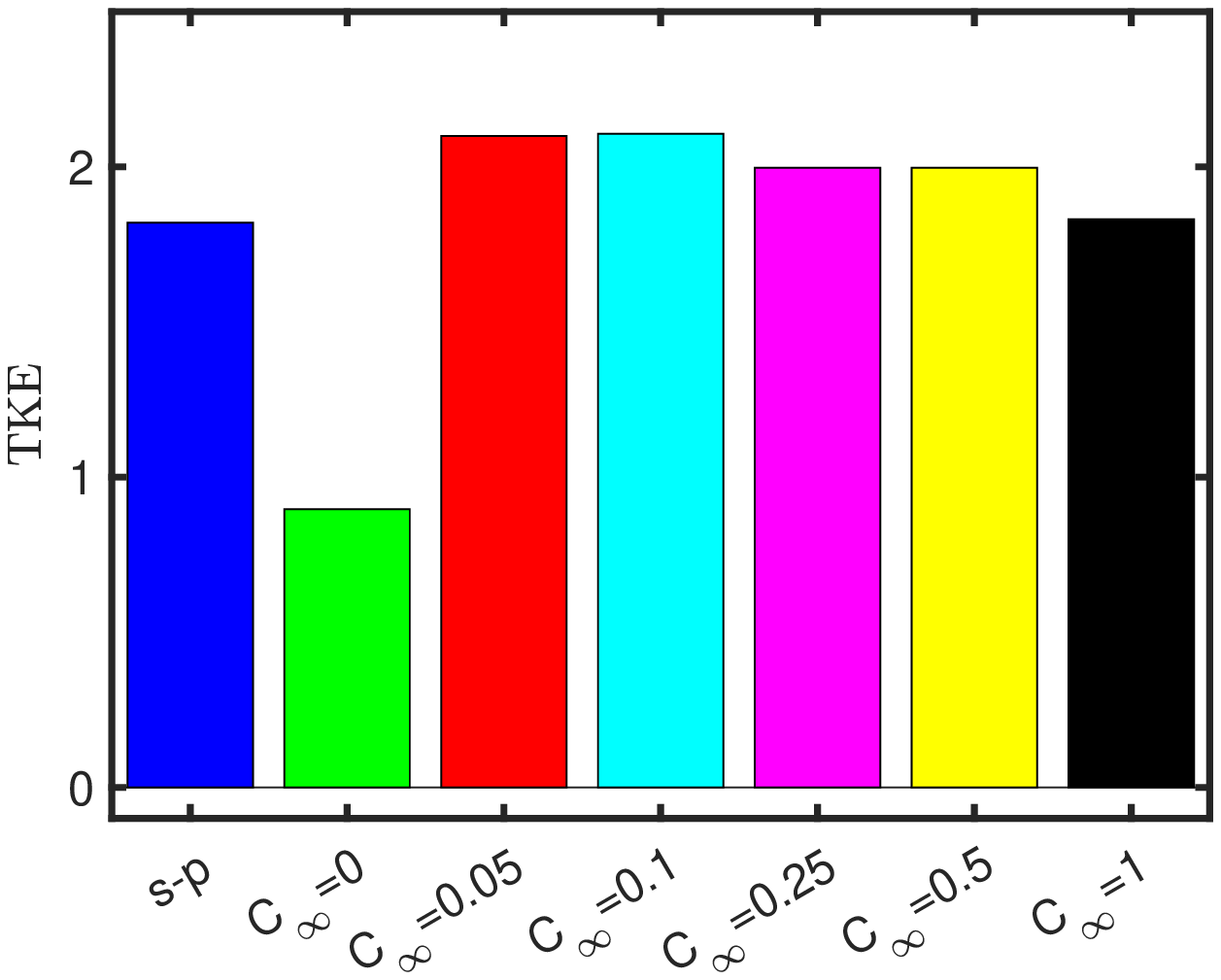}}
\caption{Newtonian Flow: (a-c) Effect of surfactant (Triton X-100) on liquid velocity fluctuations scaled by $v_\tau$ versus wall-normal distance at statistically steady state. $u^{\prime}_{rms}$,$v^{\prime}_{rms}$, and $w^{\prime}_{rms}$ are the fluctuations in the in the spanwise, streamwise and wall-normal direction, respectively. The solid lines represent the current results and the symbols represent the DNS data of \citet{vreman2014comparison}. (d) Turbulent kinetic energy (TKE) for the different bulk surfactant concentrations.}
\label{Re_stresses}
\end{figure} 

\begin{figure}
\centering
\subfloat[]{\includegraphics[width=0.5\textwidth]{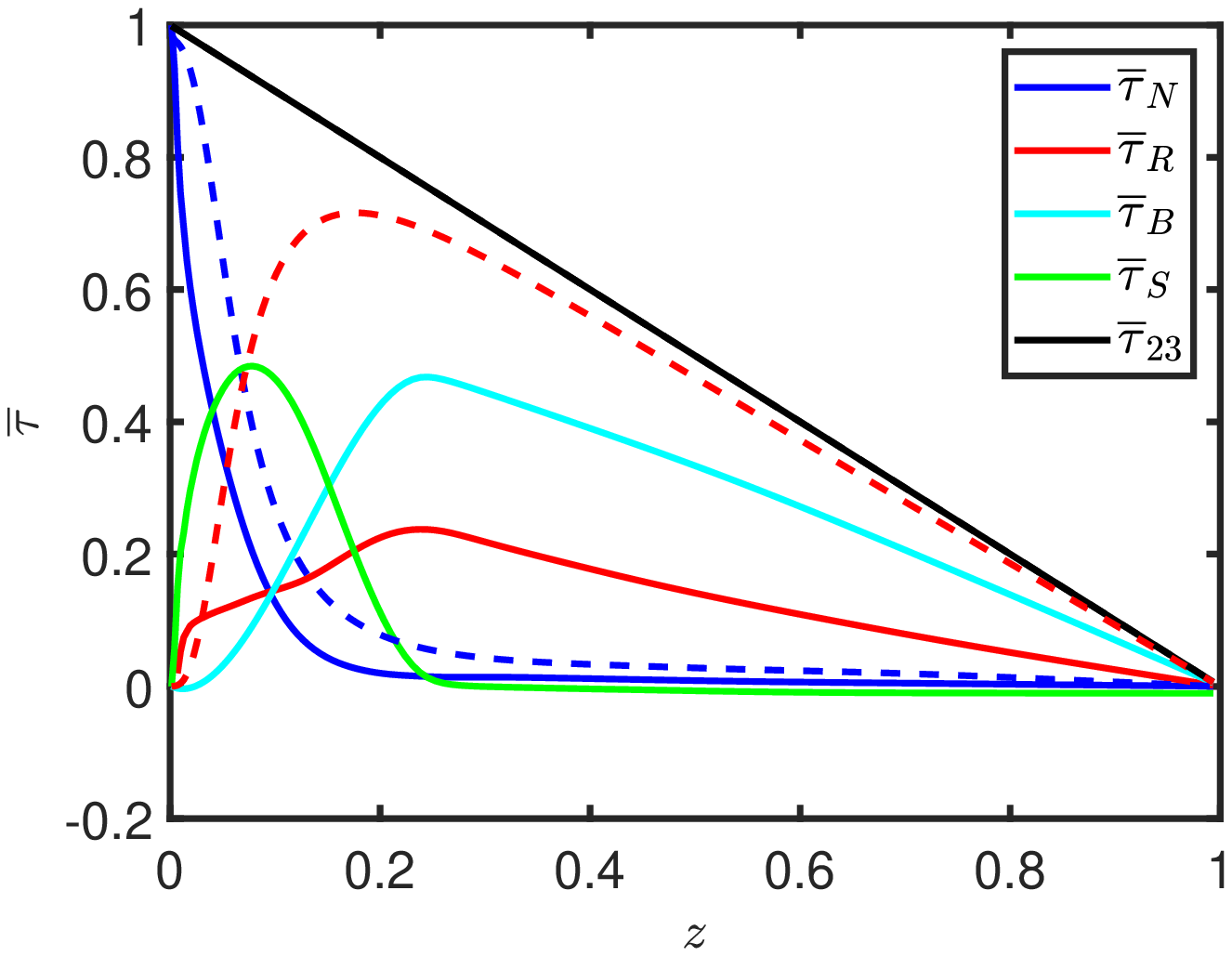}}
\subfloat[]{\includegraphics[width=0.5\textwidth]{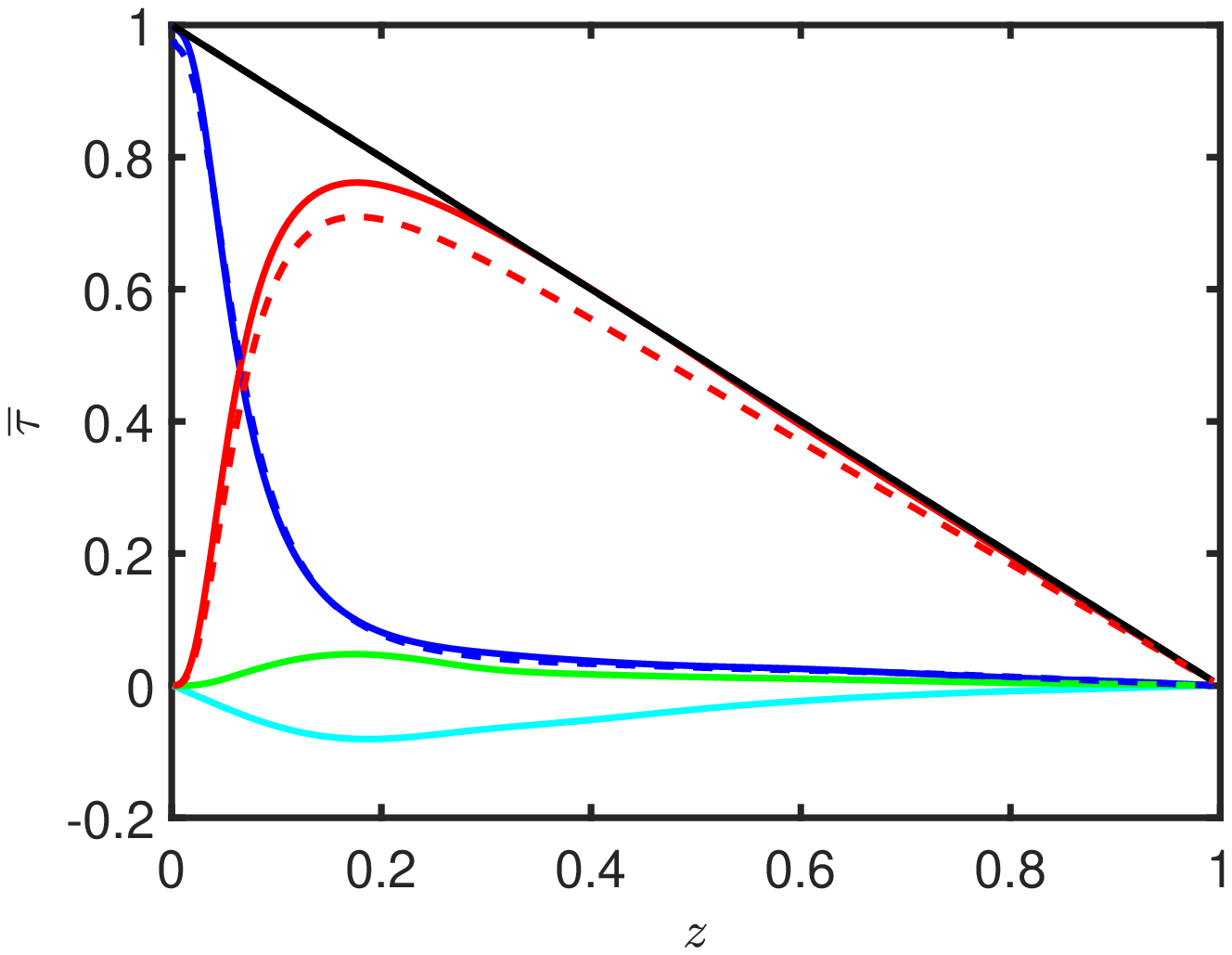}} \\ 
\subfloat[]{\includegraphics[width=0.5\textwidth]{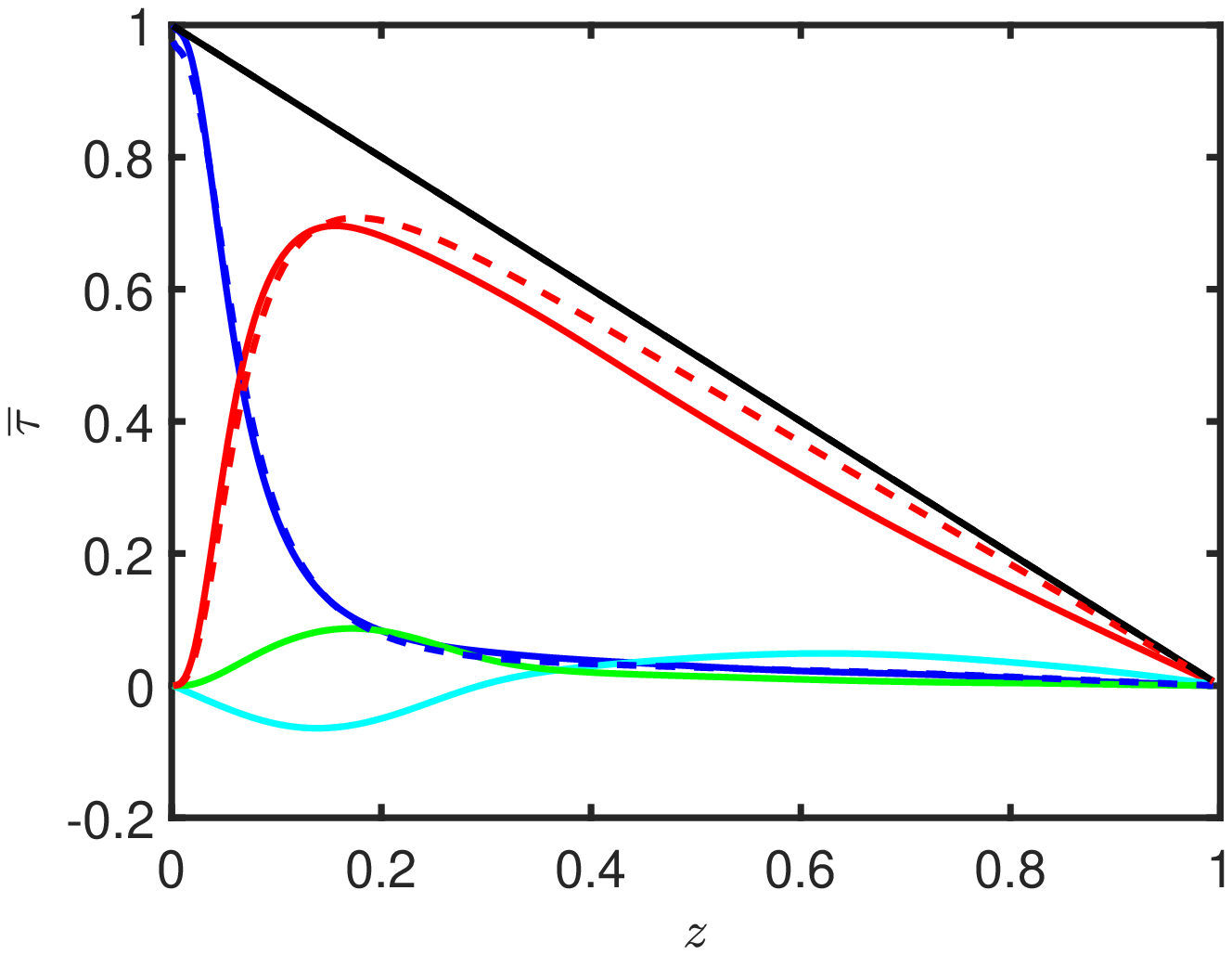}}  
\subfloat[]{\includegraphics[width=0.5\textwidth]{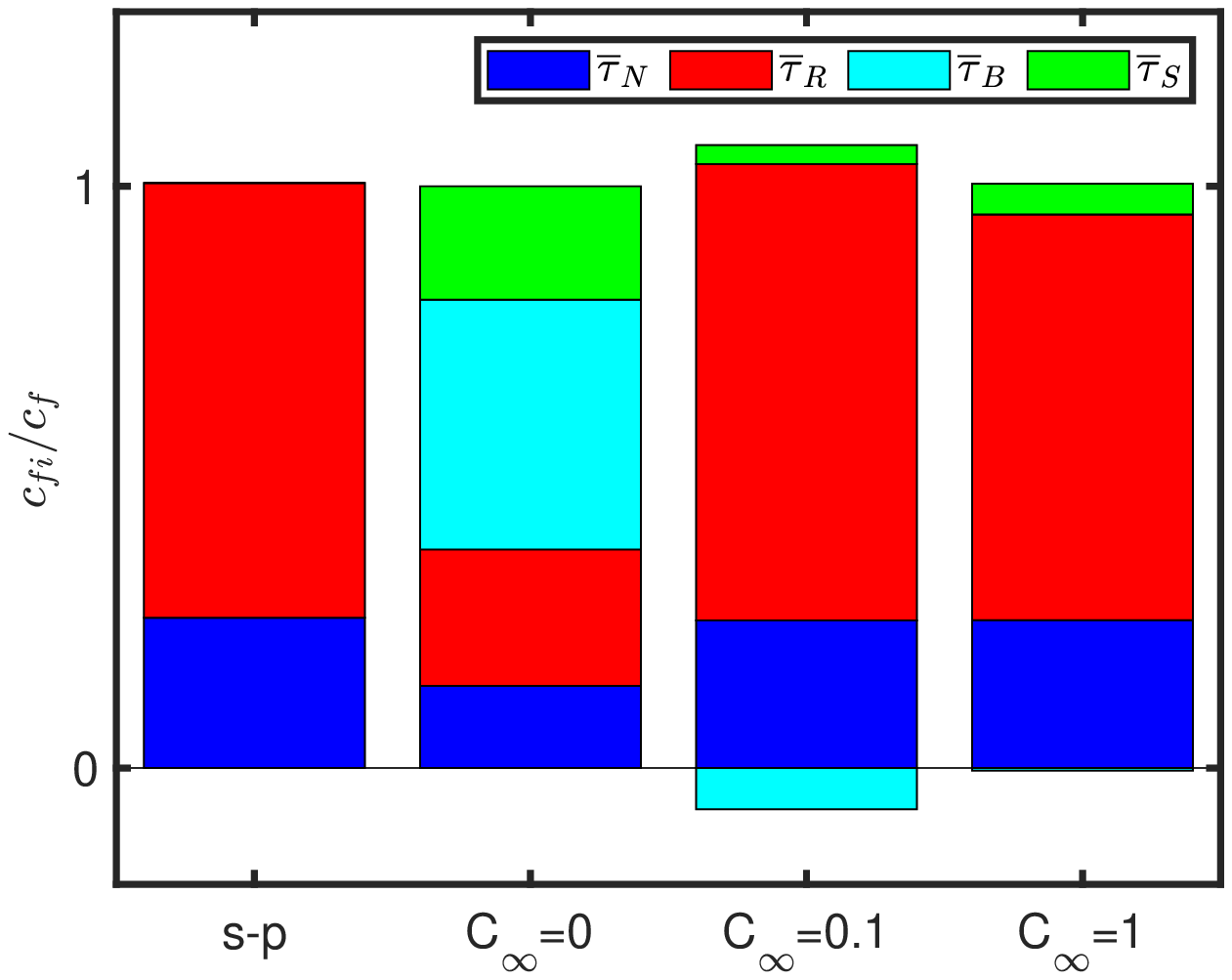}}  
\caption{Newtonian flow (Triton X-100): (a)-(c) Stress balance at a statistically steady state for $C_{\infty}=0.0$ , $C_{\infty}=0.1$ and $C_{\infty}=1.0$ respectively. Dotted lines show the corresponding single-phase stress balance. $\overline{\tau}_N$ is viscous stress , $\overline{\tau}_{R}$ is Reynolds stress, $\overline{\tau}_{B}$ is buoyancy stress, $\overline{\tau}_{S}$ is surface stress and $\overline{\tau}_{23}$ is total shear stress. All stress profiles are scaled with the wall stress $\overline{\tau}_{w}$. (d) Contribution of different stresses to the friction coefficient at various $C_{\infty}$ values.}
\label{stressBN}
\end{figure}  

The increase in drag for a turbulent bubbly flow, compared to a turbulent single-phase flow, is related to the formation of bubble-wall layers, as can be seen for the clean case ($C_\infty=0$) in Figs.~\ref{velprof} and \ref{QcritN}. The lateral migration of the bubbles to the wall is due to the inertial lift force. The aerodynamic lift force acts perpendicular to both the direction of motion and the rotation of the spherical bubble~\cite{ho_leal_1974}. The presence of a bubble wall-layer affects the liquid velocity profile, as shown in Fig.~\ref{velprof}. The velocity profile for the clean case $C_\infty=0$ does not show a clear viscous sublayer near the wall due to the presence of the bubble clusters which, as we will see, dampen the Reynolds stresses.
As shown in Fig.~\ref{QcritN}, surfactant prevents the formation of these bubble wall-layers and thus, alleviate the increase in skin friction coefficient and the mean velocity resembles those of single-phase turbulent channel flow. Clearly, flow dynamics strongly depends on the specific kinetics of surfactant  absorption, as a minute amount of Triton X-100 is sufficient to prevent the bubble wall-layers, while $100$ times more concentration of 1-Pentanol is not very effective. This is due to the amount of surfactant adsorbed on the bubble surface, which is much larger for Triton X-100 than the 1-Pentanol, as can be seen in Fig.~\ref{QcritN}. The adsorption rate of Triton X-100 is hence about $10$ times larger than that of 1-Pentanol, and the desorption rate is about $3300$ times smaller. In the case of Triton X-100, a larger amount of surfactant and its non-uniform distribution on the surface of bubble generates a strong Marangoni-induced force which prevents the motion of bubbles toward the channel wall. For $C_\infty=1$ of Triton X-100, the void fraction near the wall slightly increases, as the surfactant distribution becomes more uniform. Figure~\ref{QcritN} also shows the turbulent structures, visualized using the Q-criterion~\cite{jeong1995identification} ($Q^*/(V^*_b/(h^*))^2=0.8$), where $Q$ is the second invariant of the velocity gradient tensor. Indeed, it is clear that for $C_\infty=0$ and $C_\infty=1$ of 1-Pentanol, the turbulence is strongly attenuated, whereas for Triton X-100, the flow is in a turbulent state.

Figure~\ref{Re_stresses}(a-c) shows the root mean square of the liquid velocity fluctuations for the clean and contaminated bubbly flow (Triton X-100). For the entire $C_\infty$ range, the contaminated bubbles slightly increase the liquid velocity fluctuations in all directions with respect to the corresponding single-phase flow. This increase can be attributed to the localized disturbances caused by the wakes of the bubbles. For the clean case $(C_\infty=0)$, the fluctuations in all directions are much lower than the single-phase flow due to the significant reduction in the flow rate. However, a small peak can be seen near the wall in the spanwise direction $(u^{\prime}_{rms})$. This is due to the presence of horizontal bubble clusters in the spanwise direction. Figure~\ref{Re_stresses}(d) presents the bulk turbulent kinetic energy defined as $\mathrm{TKE}=(1/L_z)\int_{0}^{L_z} ((u^{\prime}_{rms})^2/2+(v^{\prime}_{rms})^2/2+(w^{\prime}_{rms})^2/2)\,\mathrm{d}z$ for all the cases. As expected the clean $(C_\infty=0)$ case shows a significant reduction in the TKE due to presence of the bubbles near the walls -- a reduction of $50\%$ when compared to the single-phase case, contrasting of an increase of about $15\%$ for the $C_\infty=0.05$ and $0.1$ cases. Finally, a further increase in $C_\infty$ reduces the TKE, approaching the values of the single-phase case at $C_\infty=1$.

Next, the shear stress balance is studied. Equation~\ref{NS} can be averaged over the two phases, along a wall-normal distance~\cite{pope2001turbulent}. As a result, the mean total shear stress $\overline{\tau}_{23}$ can be decomposed as 
\begin{eqnarray}
\underbrace{\int_{0}^{z}\overline{\bigg(\bigg(\frac{\rho-\rho_{av}}{Fr^2}\bigg)\mathbf{g}\bigg)}dz}_{\overline{\tau}_{B}}+\underbrace{\overline{\frac{\mu}{Re} \frac{dv}{dz}}}_{\overline{\tau}_{N}}-\underbrace{\overline{\rho v^{\prime}w^{\prime}}}_{\overline{\tau}_{R}}+ \overline{\tau}_p + \underbrace{\int_{0}^{z}\overline{\frac{1}{We} F_{\sigma}}dz}_{\overline{\tau}_{S}} = \underbrace{\overline{\tau}_w(1-z)}_{\overline{\tau}_{23}},
\label{stbN}
\end{eqnarray}
\begin{eqnarray}
F_{\sigma}=\int_A\big[\sigma(\Gamma)\kappa{\bf{n}}+\nabla_s\sigma(\Gamma)\big]\delta({\bf{x}}-{\bf{x_f}})dA,
\end{eqnarray}
where $\overline{\tau}_w$ is the average wall shear stress. From left to right, the terms are the buoyancy stress ($\overline{\tau}_{B}$), the viscous stress ($\overline{\tau}_N$), the Reynolds stress ($\overline{\tau}_{R}$) and the surface stress ($\overline{\tau}_{S}$). Note that $\overline{\tau}_{p}$ is zero for a Newtonian flow. Figure~\ref{stressBN}(a-c) shows the stress balance for the clean and contaminated bubbly flows at a statistically steady state. In Figure~\ref{stressBN}(d) we present contribution of each stress $\overline{\tau}_{i}$ to the skin friction coefficient $c_f$ using the FIK (Fukagata, Iwamoto $\&$ Kasagi) identity \cite{Fukagataetal2002, yu_xia_guo_lin_2021}. The normalized contribution can be expressed as $c_{fi}=\int_{0}^{1} 6 (1-z) \overline{\tau}_{i} \,\mathrm{d}z$. For $C_\infty=0$, the formation of bubble clusters near the wall completely alters the stress balance compared to the single-phase flow. Due to the high void fraction near the wall, the applied pressure gradient is balanced by the weight of the mixture in the middle of the channel. The Reynolds and viscous stresses decrease resulting in enhancement of buoyancy and surface stresses. Indeed, Figure~\ref{stressBN}(d) clearly shows that the relative contribution of $\overline{\tau}_{R}$ to $c_f$ decreases from $74\%$ for the single-phase flow to $23\%$ for the clean case ($C_\infty=0$), consistently with the observed turbulence attenuation in the presence of wall layers. For $C_\infty=0.1$, the stress balance of contaminated bubbly flow becomes similar to the single-phase flow, i.e., the major contributors to the total shear stress are viscous and Reynolds stresses. Reynolds stress is even slightly higher than that of the single-phase flow but for $C_\infty=1.0$, it slightly decreases due to the increase in the void fraction near the wall. Surface stress for both $C_\infty=0.1$ and $1.0$ become nearly zero and the buoyancy term becomes slightly negative near the wall due to very low void fraction and elimination of bubble clusters.

\subsection{Combined effects: Viscoelasticity and Surfactants}
Next, simulations are performed to examine the combined effects of viscoelasticity and surfactant (Triton X-100) on dynamics of bubbly flow. For this purpose, single-phase viscoelastic turbulent flow is first simulated for $\mathrm{Wi}=4$ and $\mathrm{Wi}=8$. Then, we add randomly positioned contaminated bubbles at fixed volume fraction of 3$\%$ to the viscoelastic base flow. For contaminated bubbles, simulations are performed varying the surfactant concentration $C_\infty= 0.1,0.25$ and $0.5$, at a fixed Weissenberg number ($\mathrm{Wi}=4$). After that, the Weissenberg number is changed from $\mathrm{Wi}= 4$ to $\mathrm{Wi}=8$, while keeping $C_\infty= 0.5$.

\begin{figure}
\centering
\subfloat[]{\includegraphics[width=0.48\textwidth]{4a-Figure.eps}}
\subfloat[]{\includegraphics[width=0.48\textwidth]{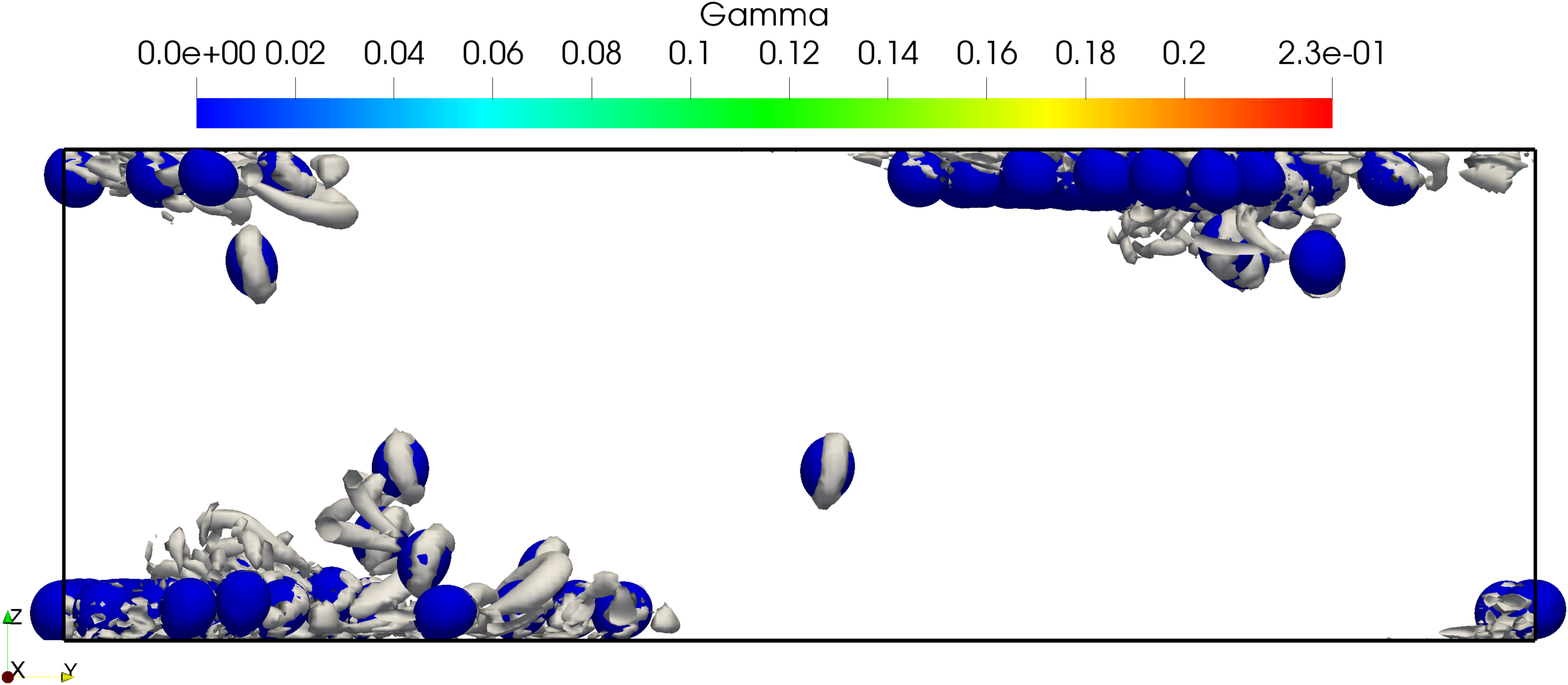}}\\
\subfloat[]{\includegraphics[width=0.48\textwidth]{4b-Figure.eps}}
\subfloat[]{\includegraphics[width=0.48\textwidth]{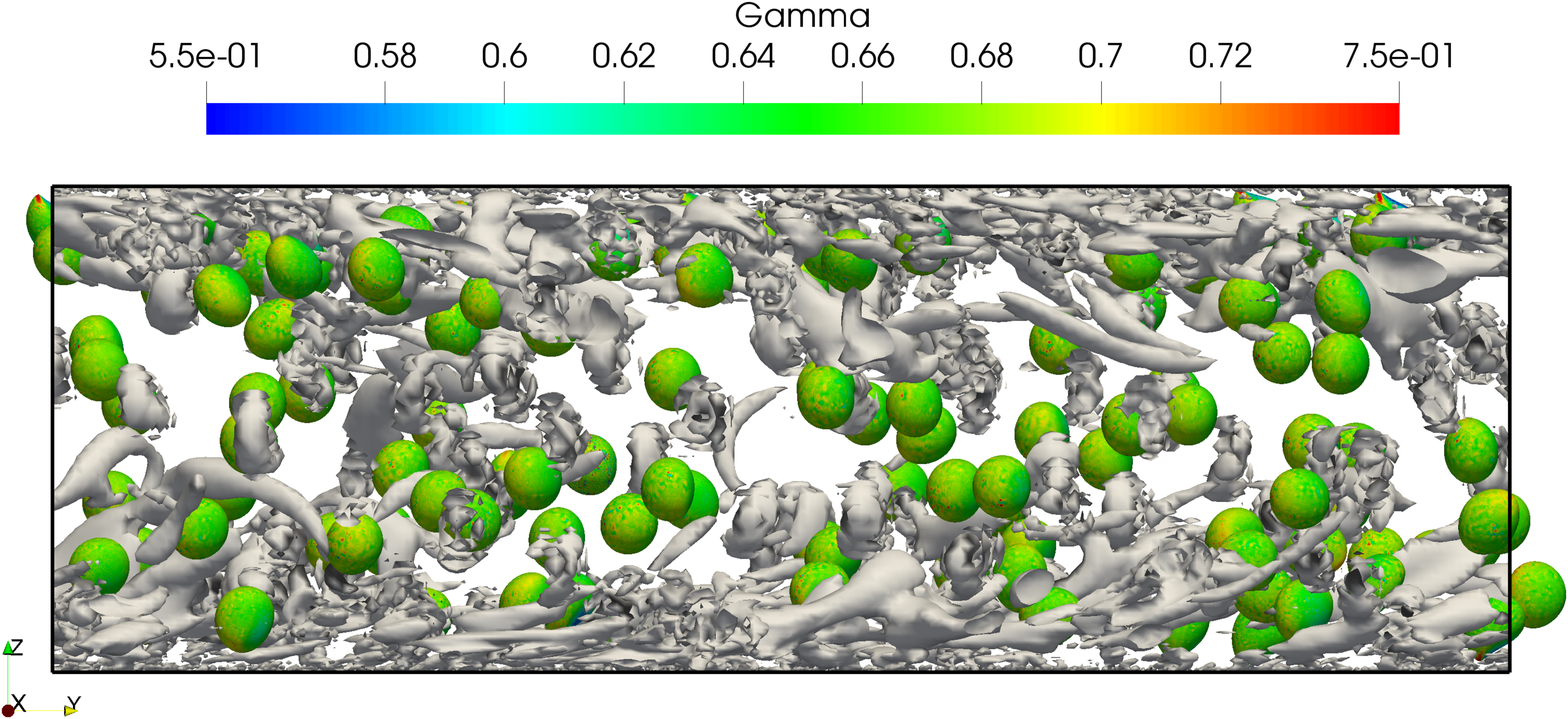}}\\
\subfloat[]{\includegraphics[width=0.48\textwidth]{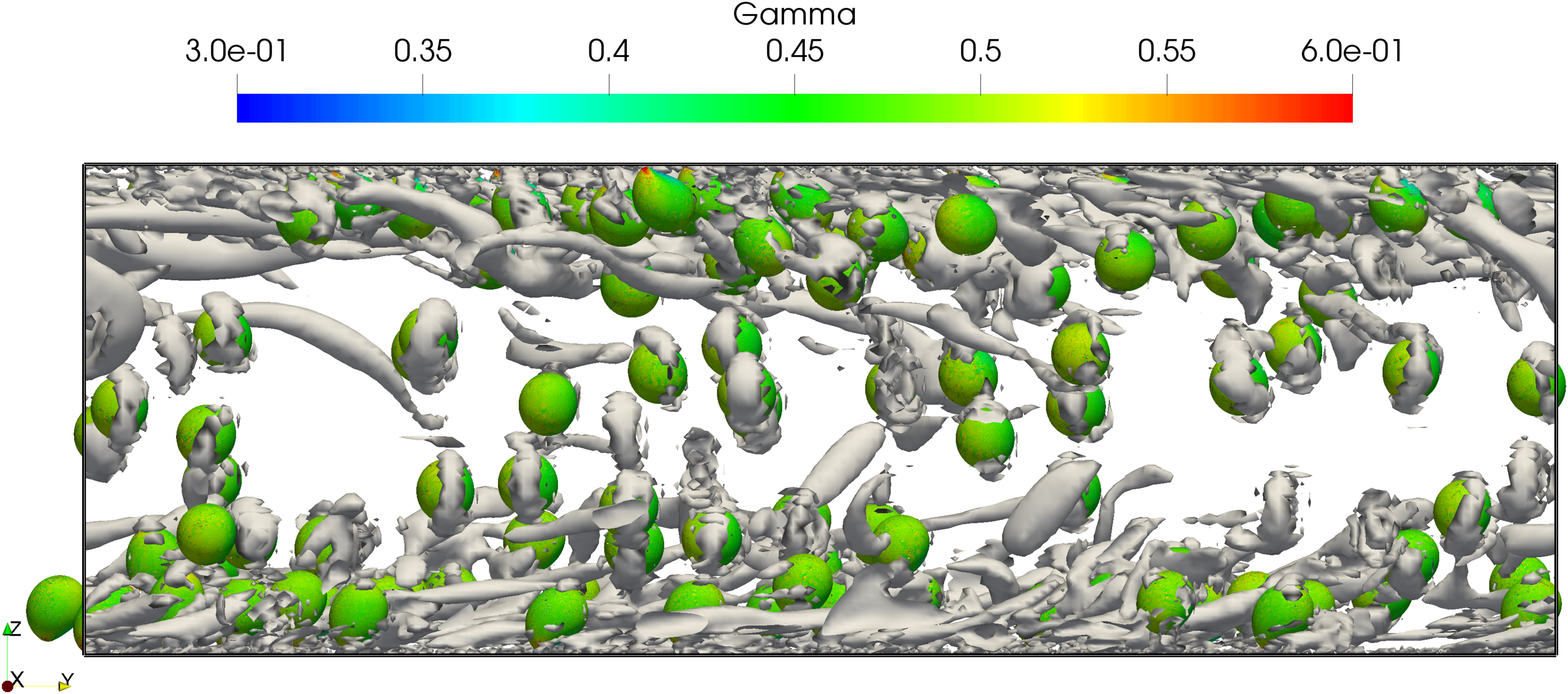}}
\subfloat[]{\includegraphics[width=0.48\textwidth]{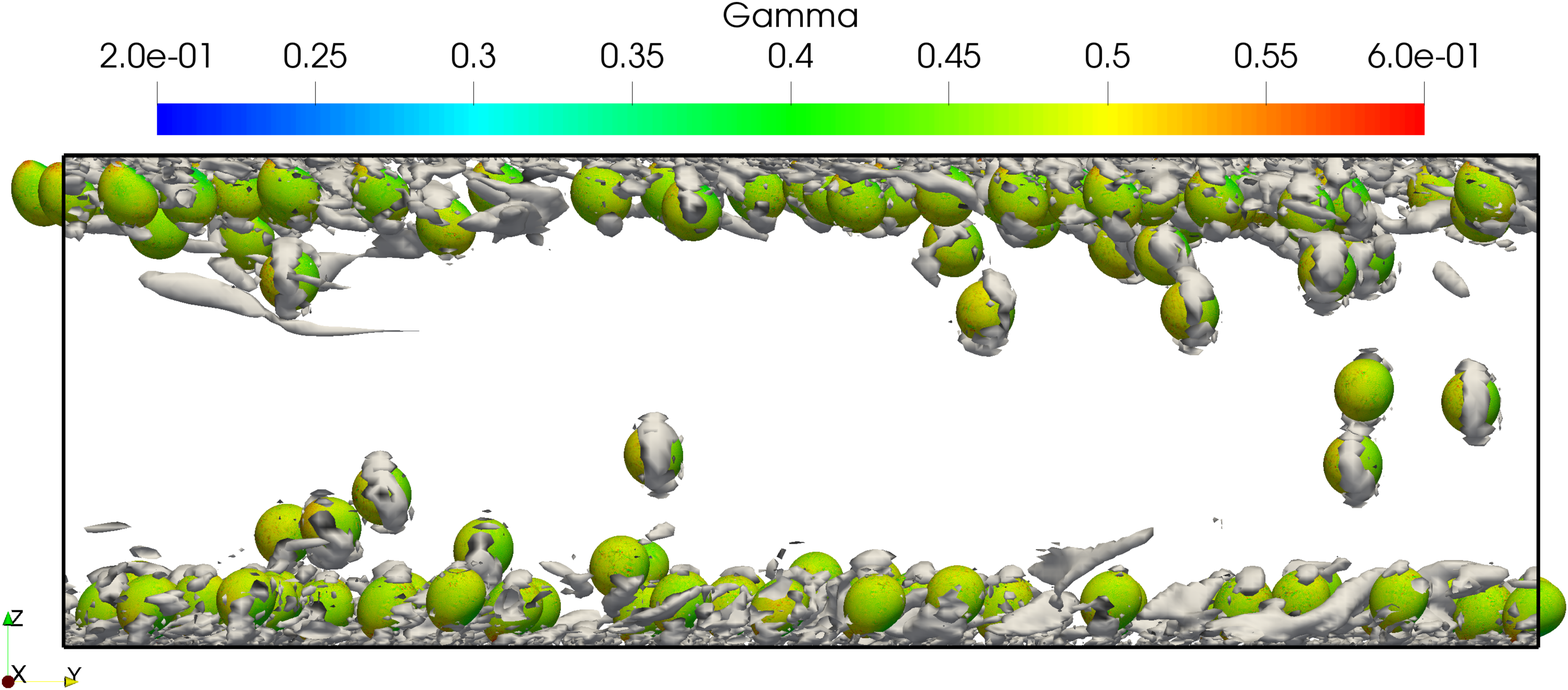}}\\
\caption{The vortical structures at the statistically steady state. (a) $C_{\infty}=0.0,\mathrm{Wi}=0$ (b) $C_{\infty}=0.0$,$\mathrm{Wi}=4$ (c) $C_{\infty}=0.1,\mathrm{Wi}=0$ (d) $C_{\infty}=0.1$,$\mathrm{Wi}=4$ (e) $C_{\infty}=0.5$,$\mathrm{Wi}=4$ (f) $C_{\infty}=0.5$,$\mathrm{Wi}=8$. The vortical structures are visualized using $Q$-criterion~\cite{jeong1995identification}, i.e., the iso-contours of the normalized second invariant of the velocity-gradient tensor, $Q^*/(V^*_b/(h^*))^2=0.8$, are plotted. The contours on the bubble surface represent interfacial surfactant concentration $(\Gamma)$.}
\label{Qcrit_V}
\end{figure}

\begin{figure}
\centering
\includegraphics[width=0.49\textwidth]{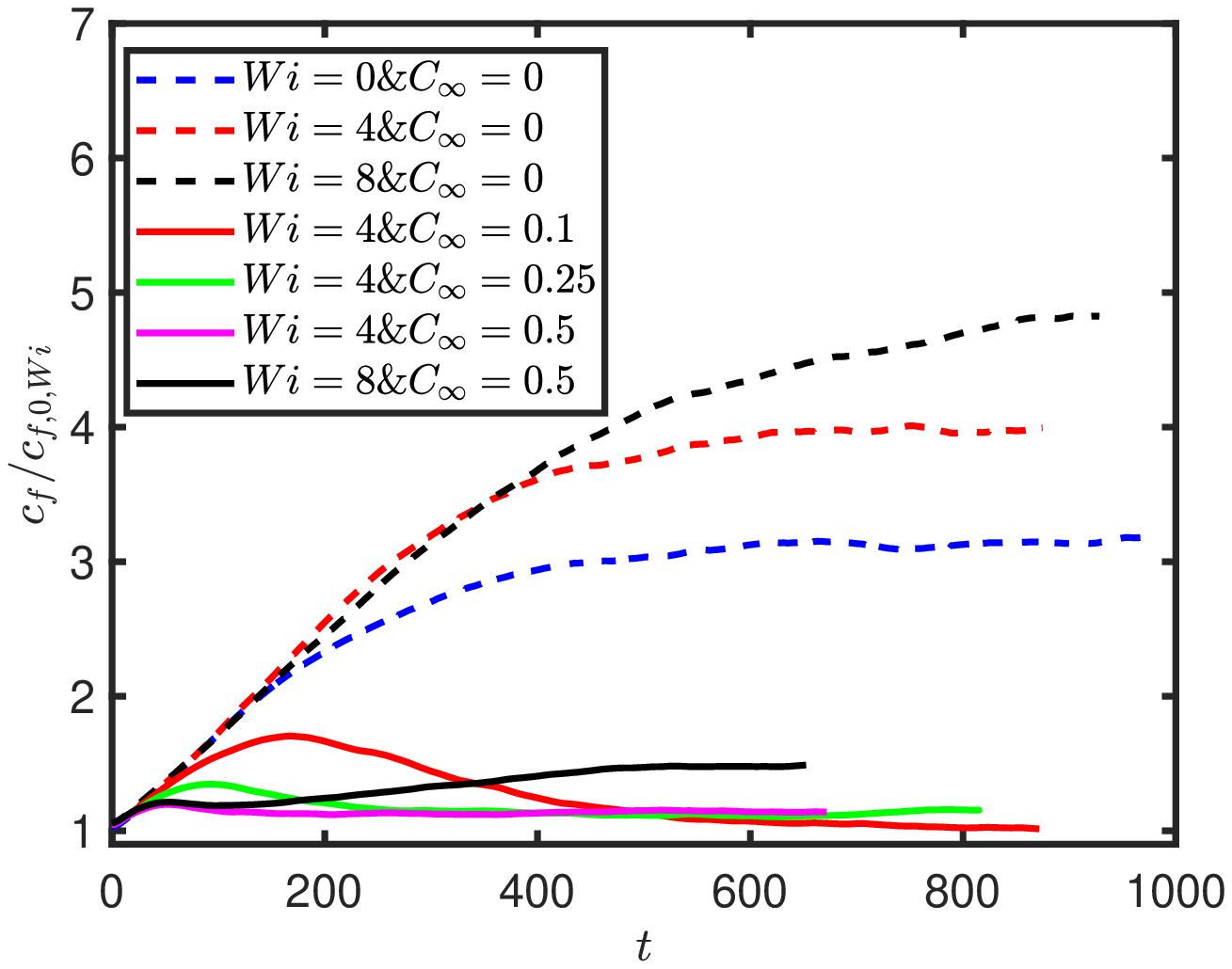}
\includegraphics[width=0.49\textwidth]{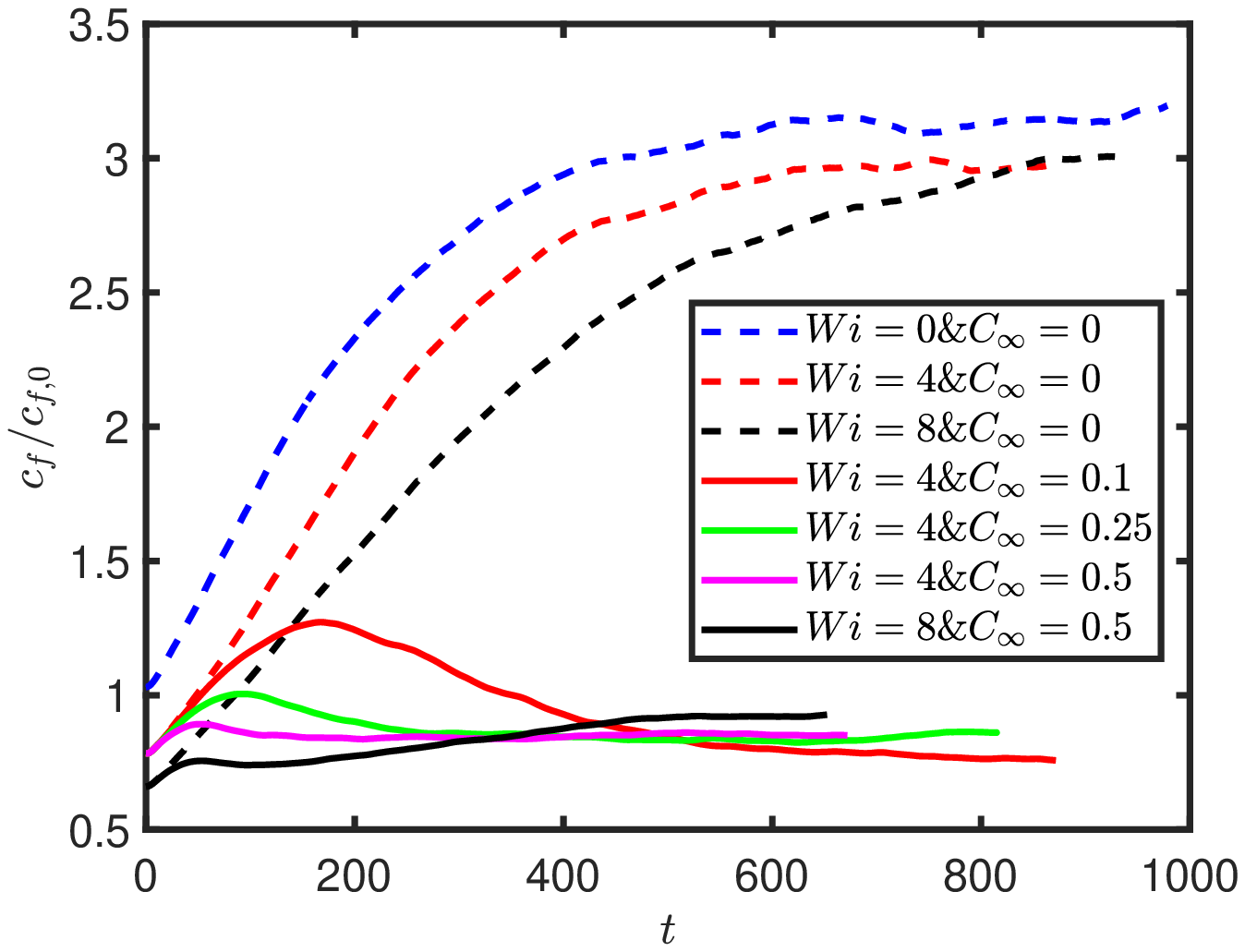}
\caption{Skin friction coefficient $(c_f)$ of the viscoelastic bubbly flow normalized by the value of the corresponding single-phase flow (left panel) and by the value of the Newtonian single-phase flow (right panel) at a statistically steady state.}
\label{SFCV}
\end{figure}

\begin{figure}
\centering
\subfloat[]{\includegraphics[width=0.48\textwidth]{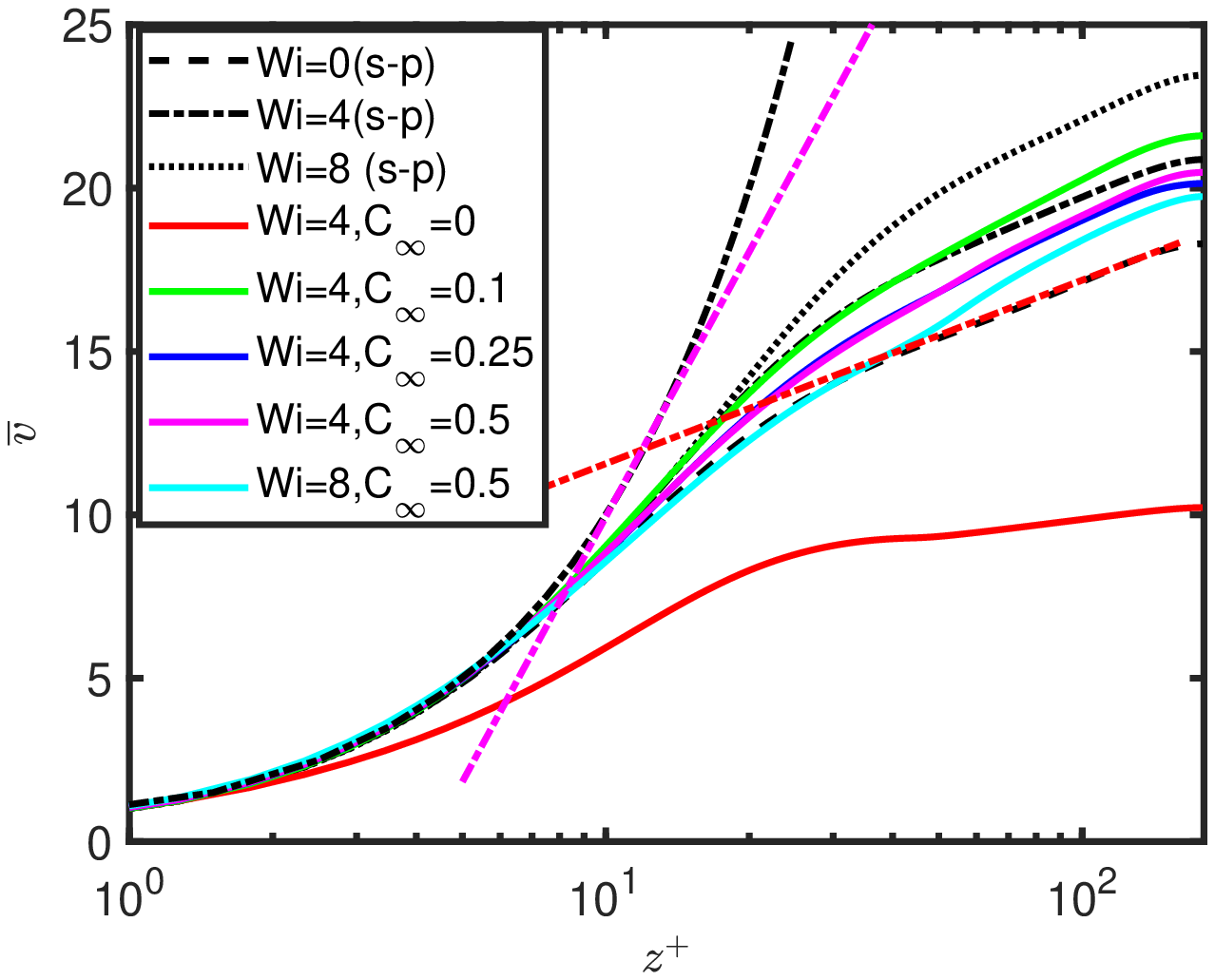}}
\subfloat[]{\includegraphics[width=0.48\textwidth]{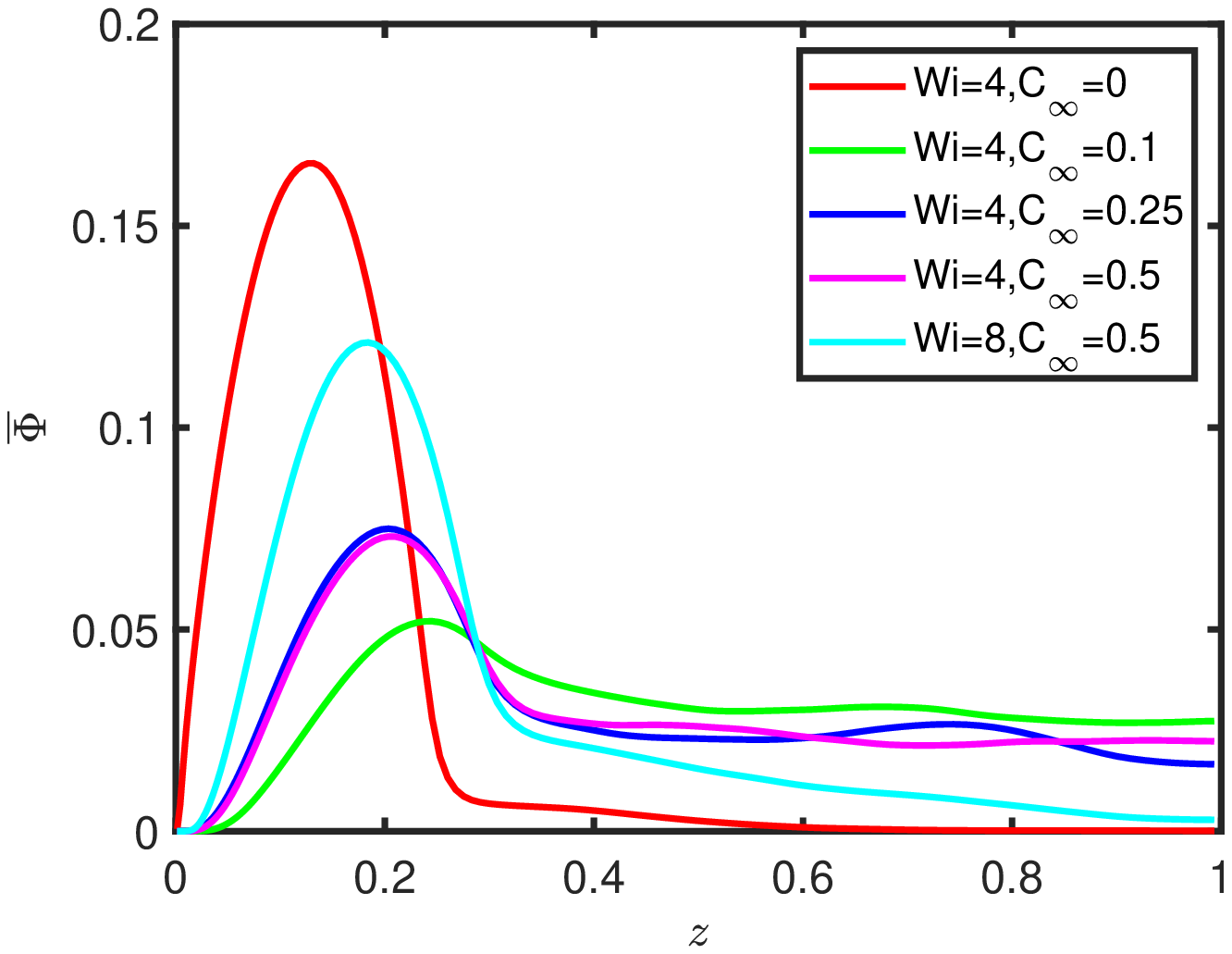}}\\
\subfloat[]{\includegraphics[width=0.48\textwidth]{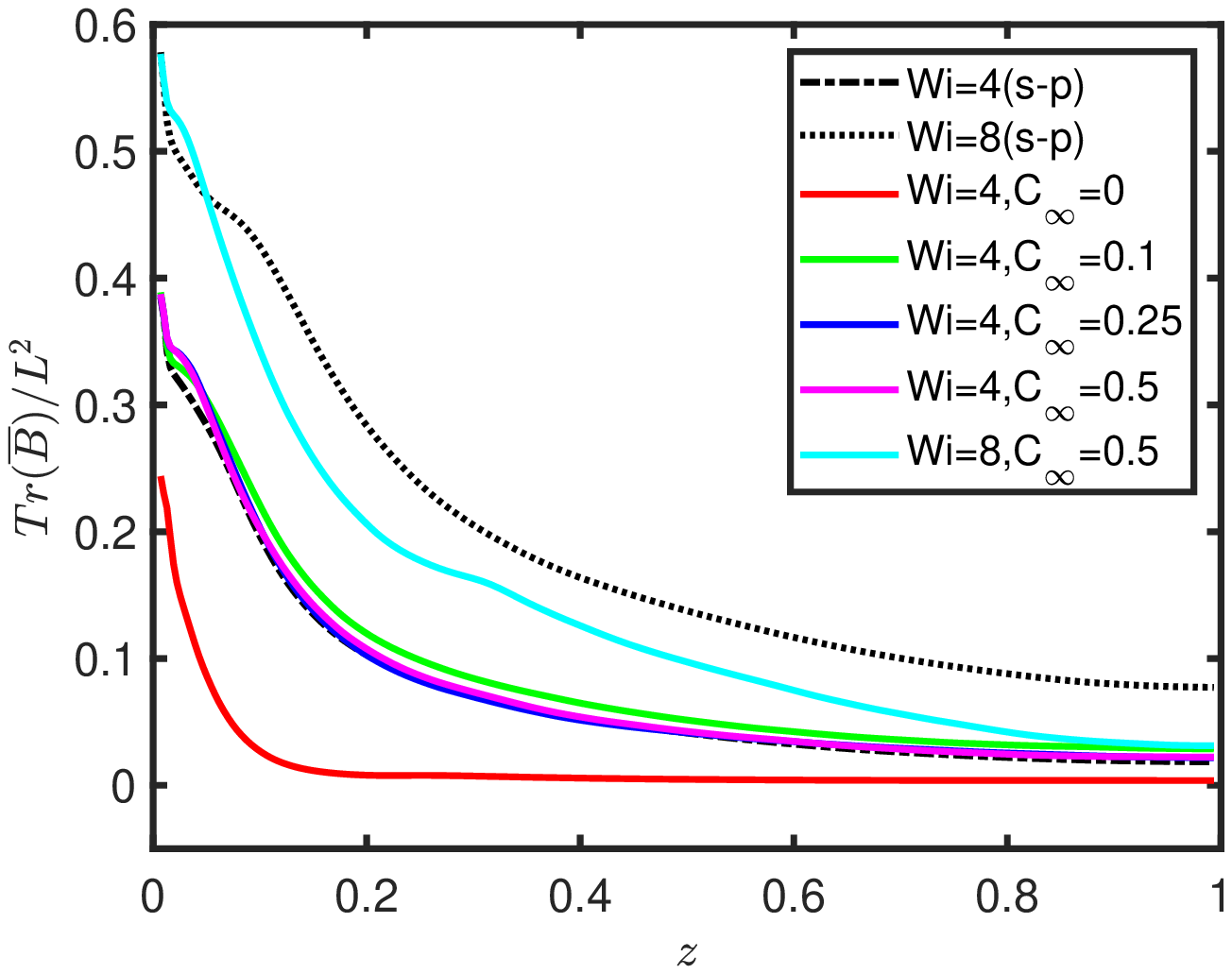}}
\caption{The statistically steady state mean profiles of (a) the mean velocity, (b) the void fraction and (c) the polymer stretching, plotted against wall-normal distance. In (a), the computed inner-scaled mean velocity profiles are compared with the law of the wall, i.e., $v^+ = z^+$ in the laminar sublayer (black dash-dotted line) and $v^+ = 2.5 \ln z^+ + 5.5$ in the logarithmic region (red dash-dotted line), as well as the MDR asymptote, $v^+ =11.7 \ln z^+ -17$ (magenta dash-dotted line).}
\label{statsV}
\end{figure}

\begin{figure}
\centering
\subfloat[]{\includegraphics[width=0.48\textwidth]{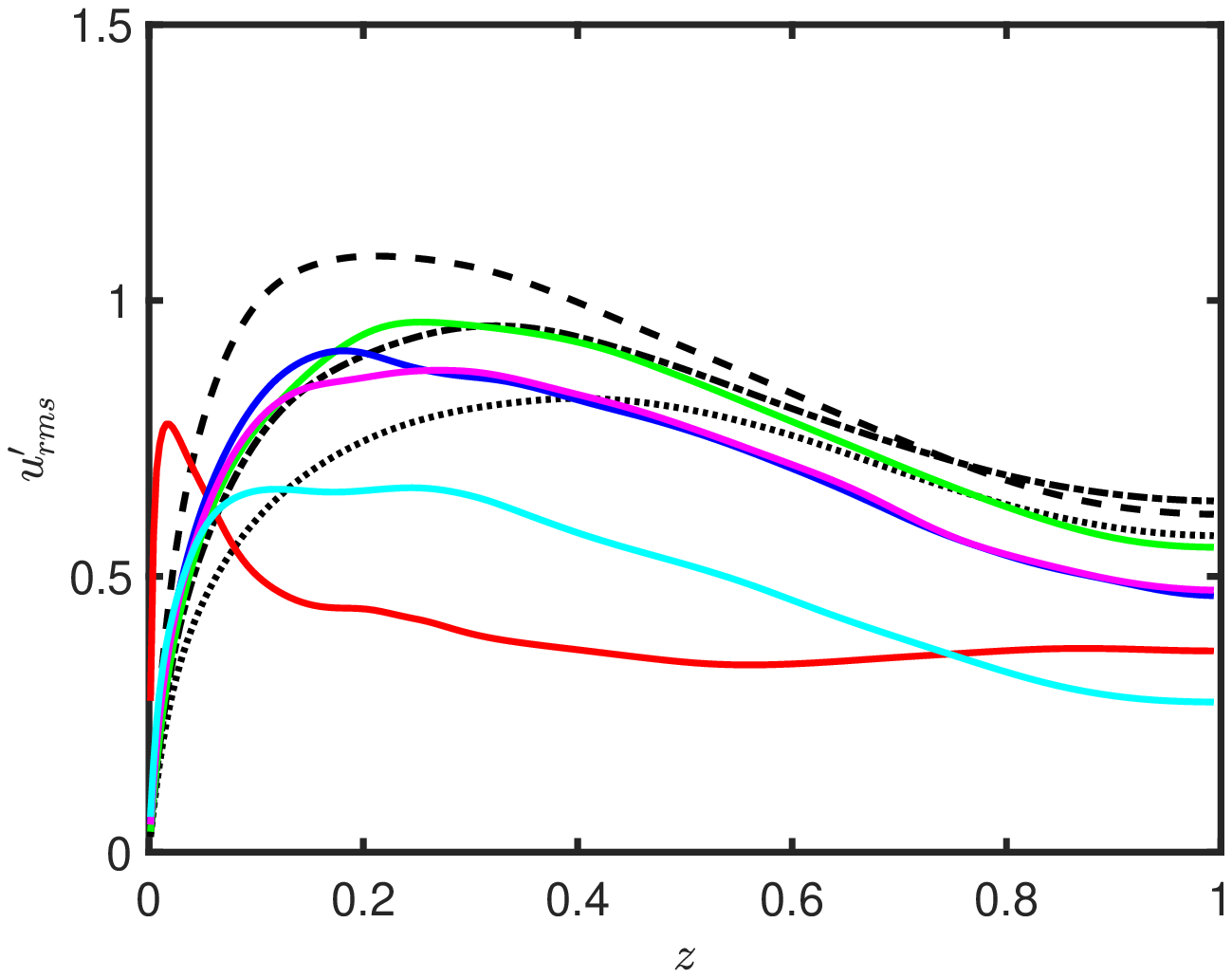}}
\subfloat[]{\includegraphics[width=0.48\textwidth]{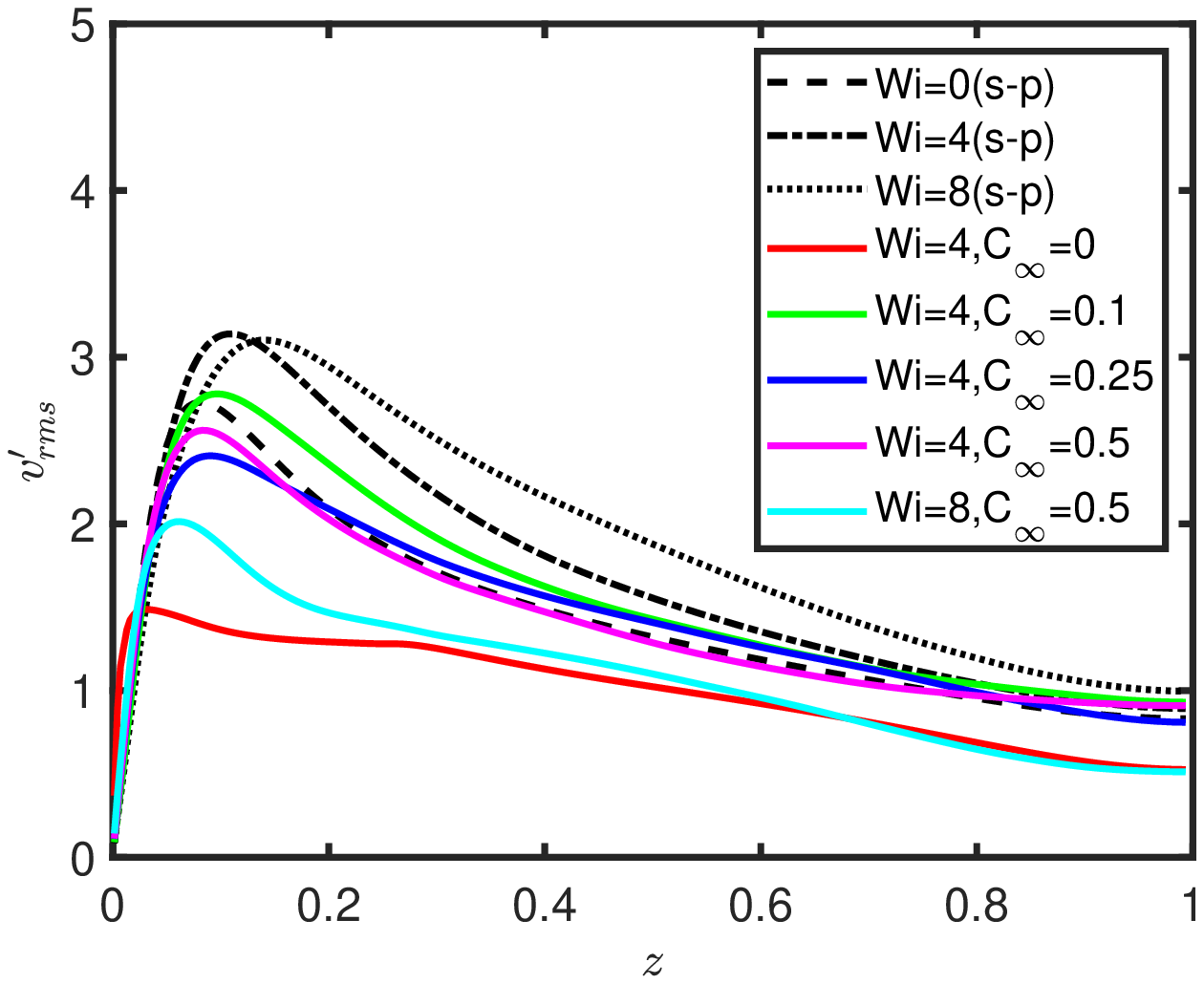}}\\
\subfloat[]{\includegraphics[width=0.48\textwidth]{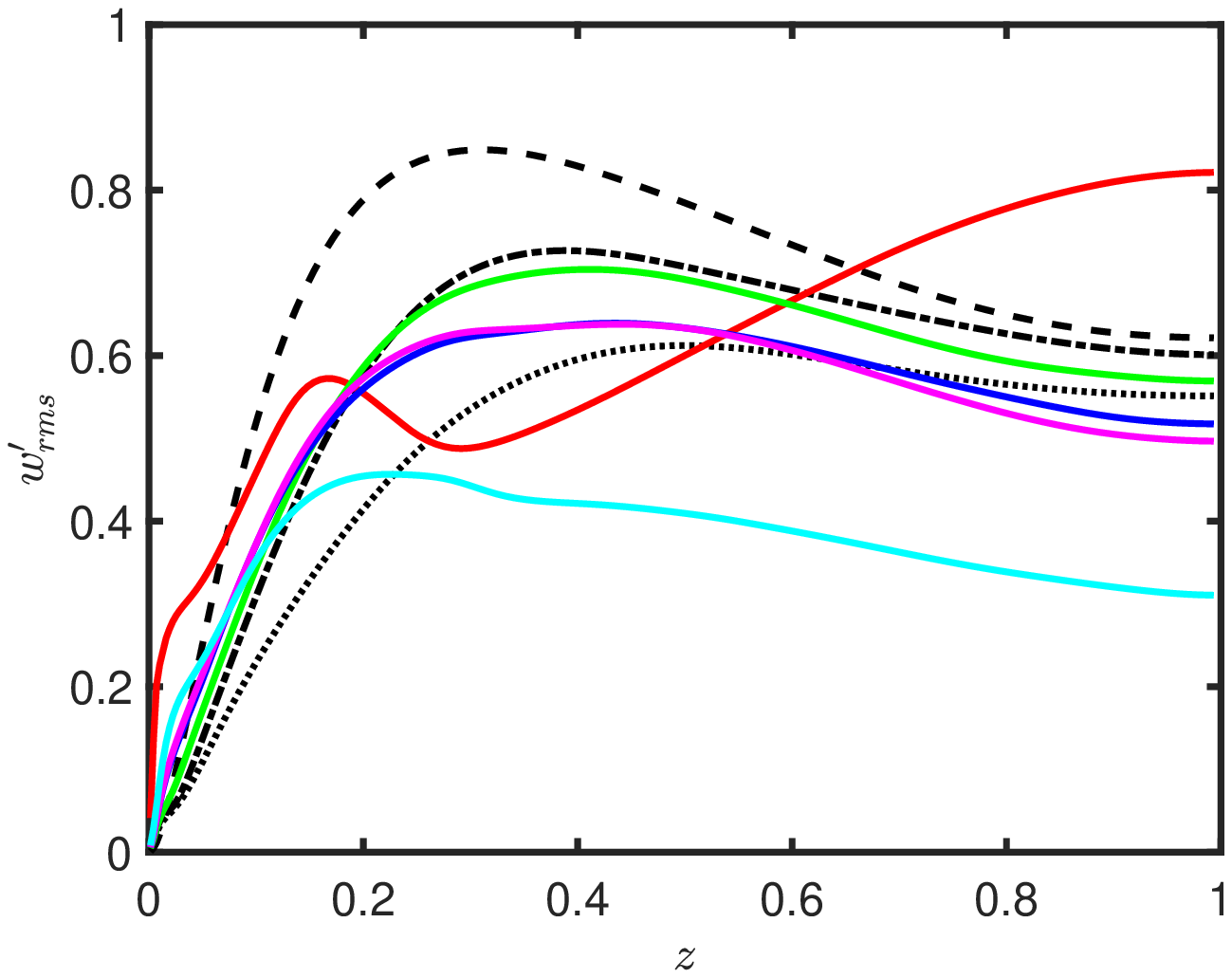}}
\subfloat[]{\includegraphics[width=0.48\textwidth]{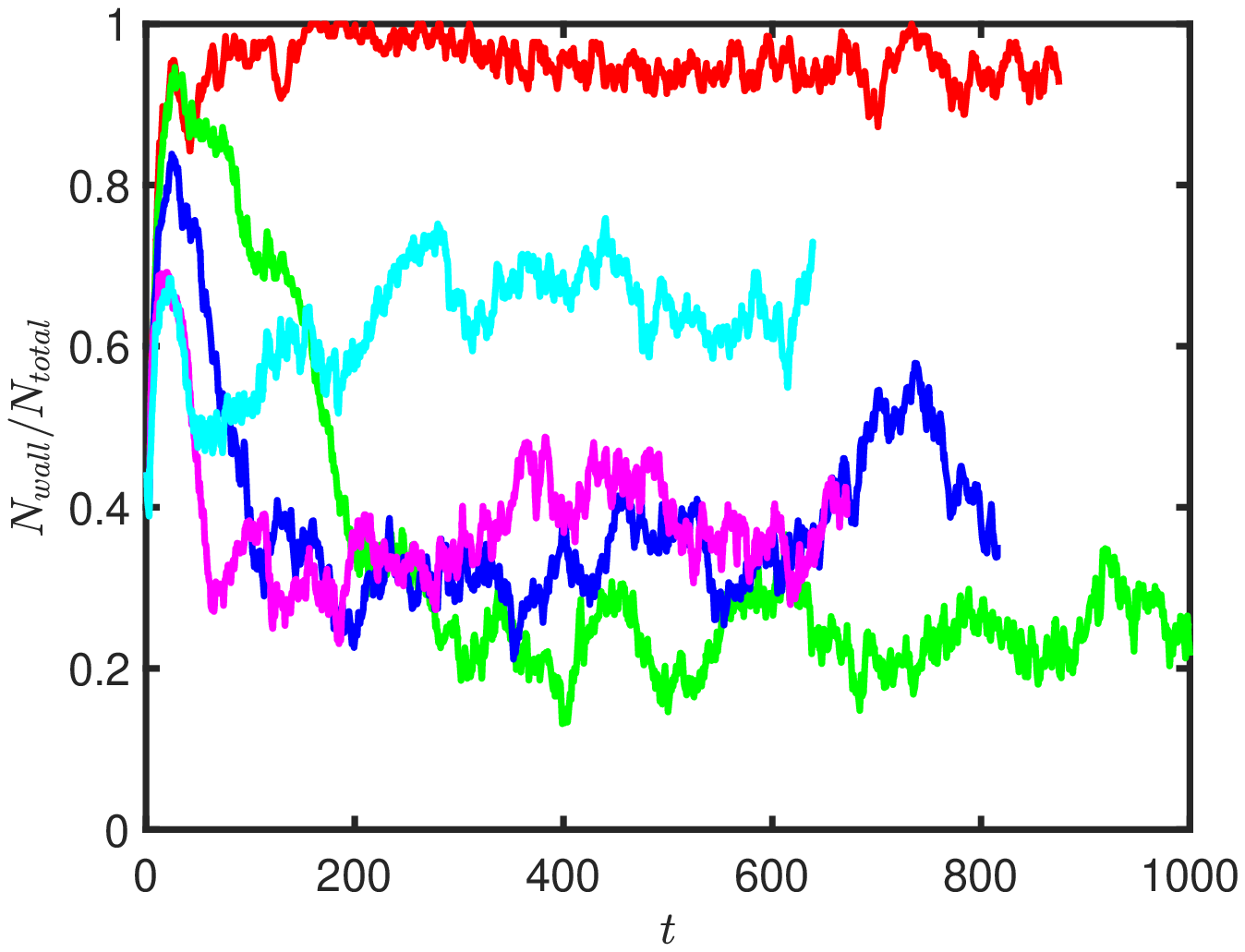}}
\caption{(a)-(c) Liquid velocity fluctuations scaled by $v_\tau$ versus wall-normal distance at statistically steady state. $u^{\prime}_{rms}$,$v^{\prime}_{rms}$, and $w^{\prime}_{rms}$ are the root mean square of the velocity fluctuations in the spanwise, streamwise and wall-normal direction, respectively. (d) Transient result: $N_{wall}/N_{total}$ is the number of bubbles in the bubble wall-layer. The width of the bubble wall-layer is defined as a diameter of bubble.}
\label{RES_V}
\end{figure}

\begin{figure}
\centering
\subfloat[]{\includegraphics[width=0.48\textwidth]{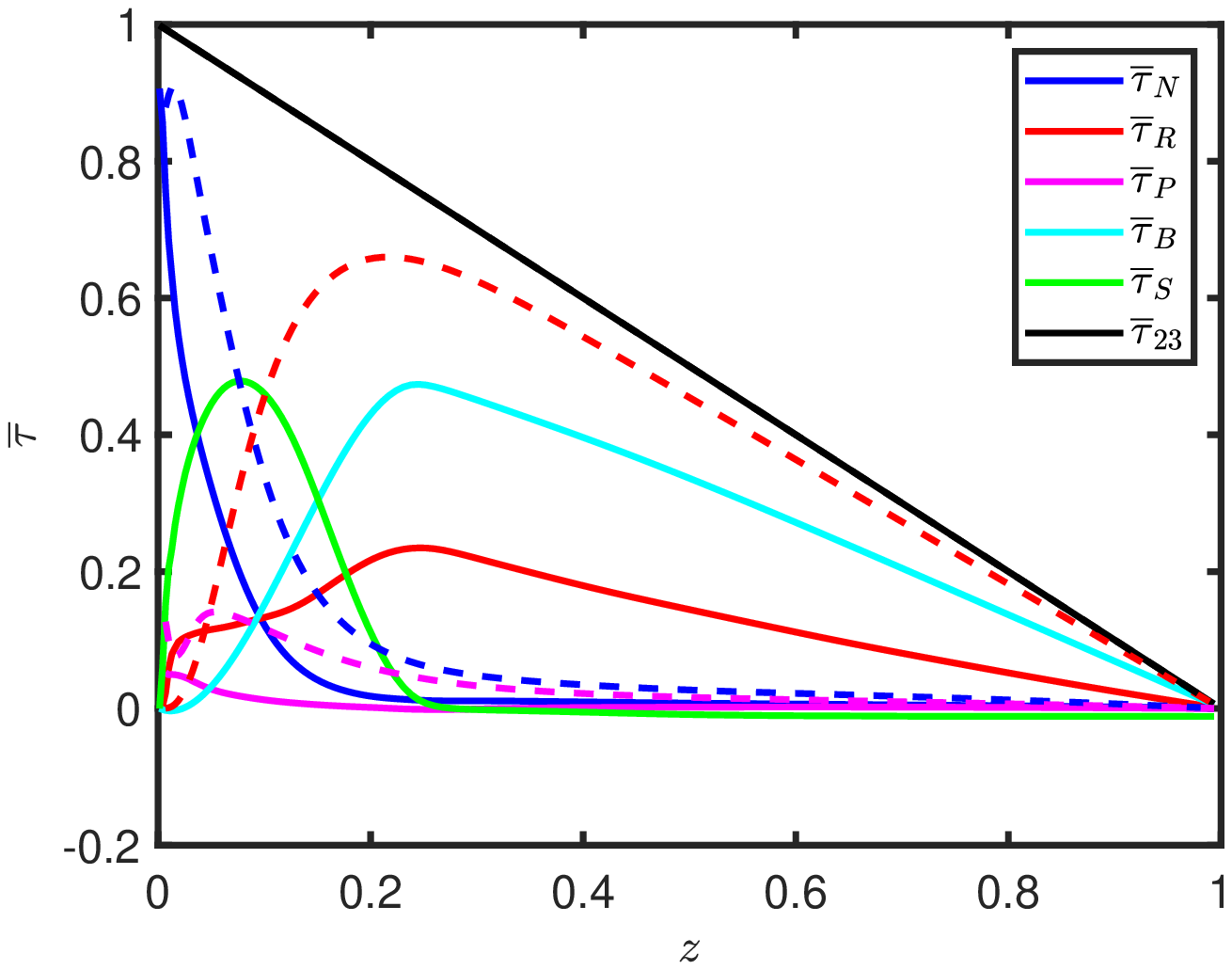}}
\subfloat[]{\includegraphics[width=0.48\textwidth]{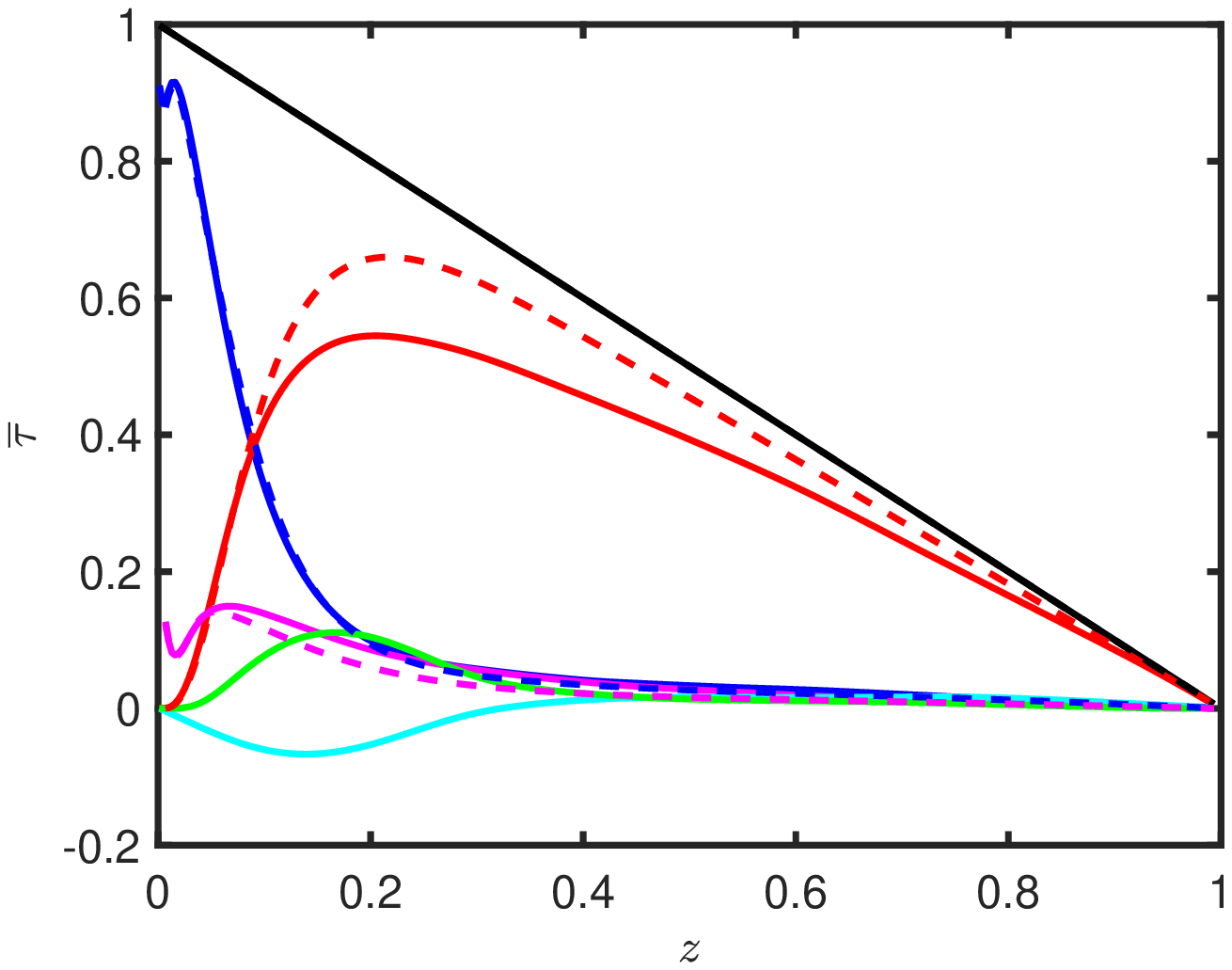}}\\
\subfloat[]{\includegraphics[width=0.48\textwidth]{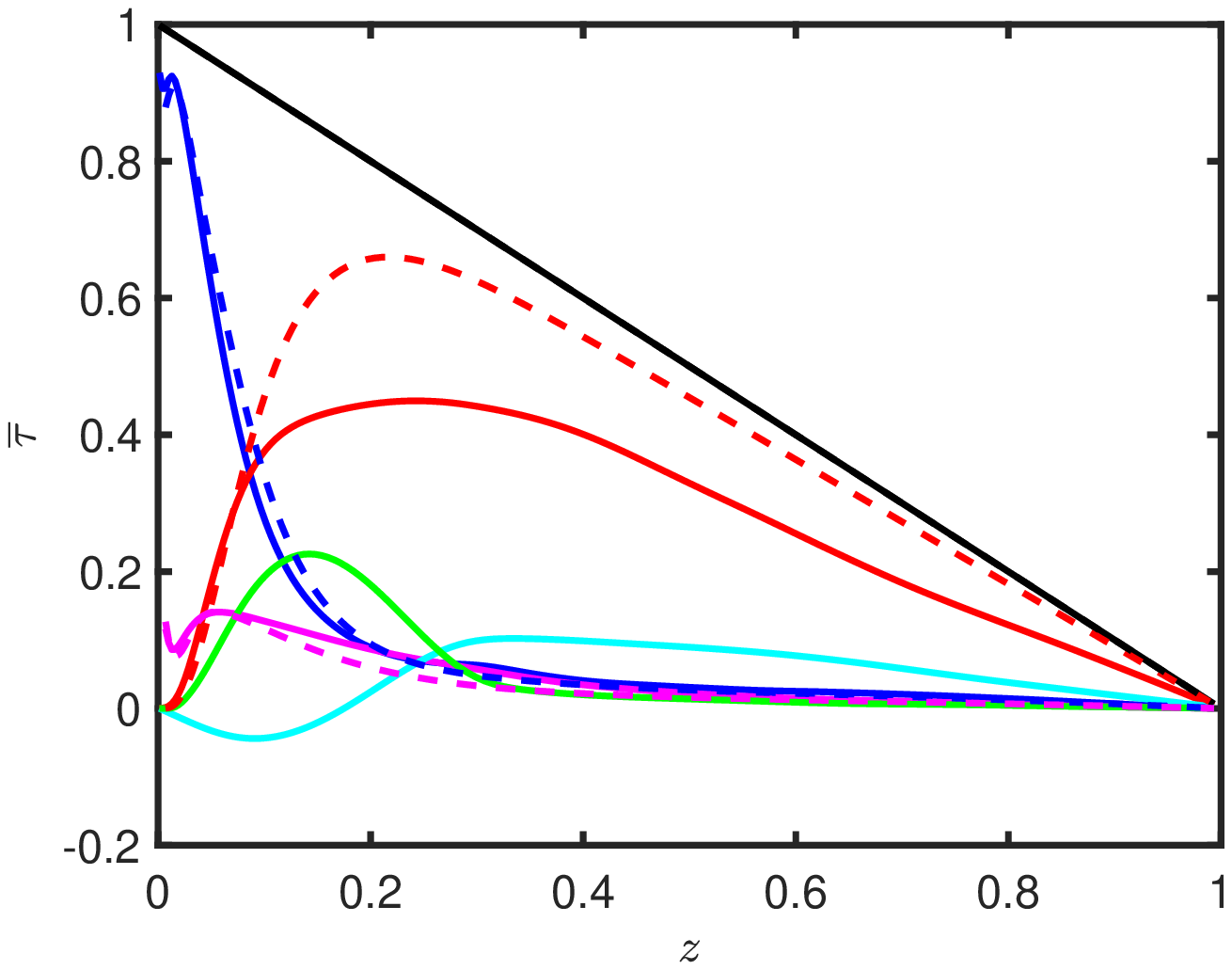}}
\subfloat[]{\includegraphics[width=0.48\textwidth]{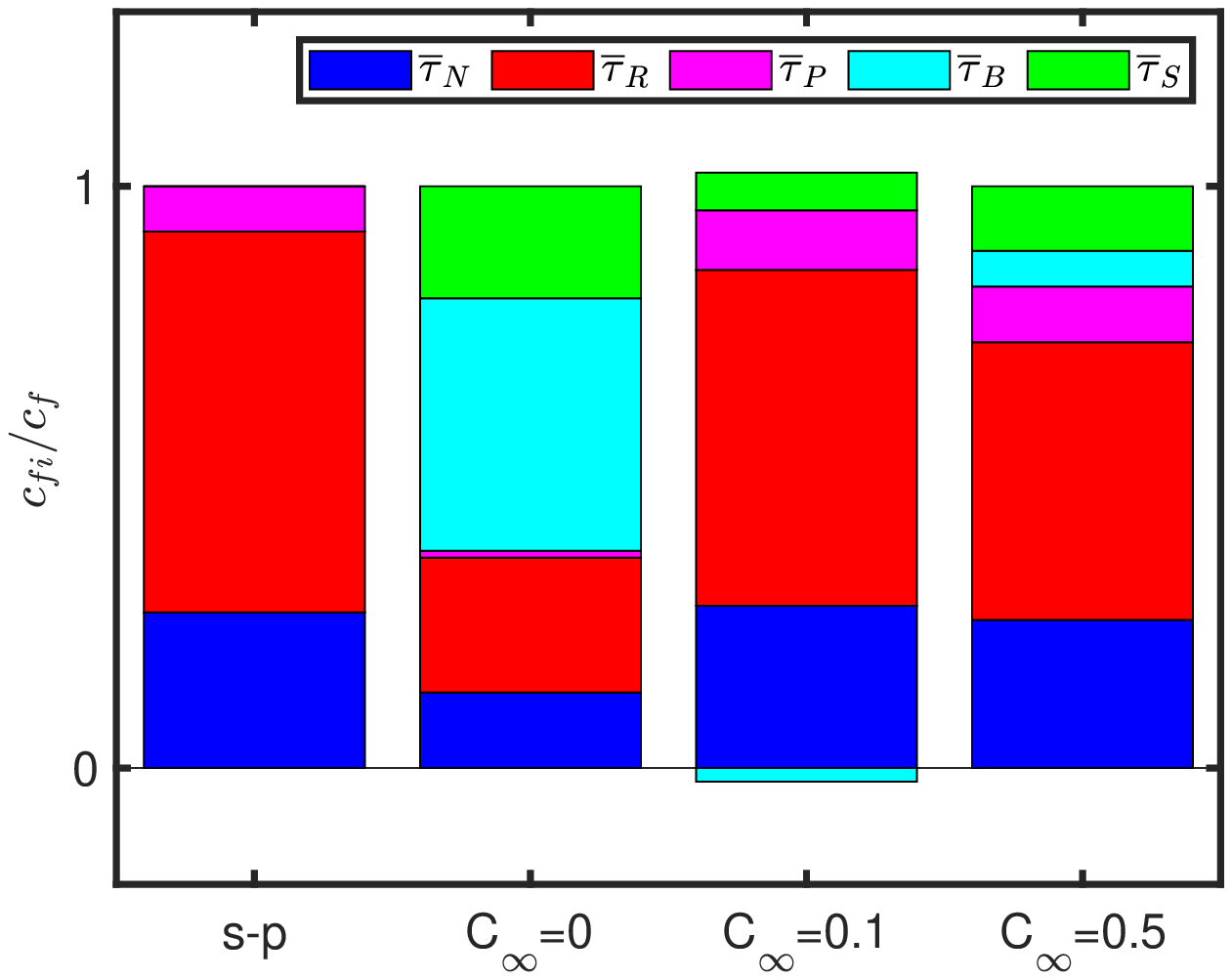}}

\caption{(a-c) Stress balance at the statistically steady state for $\mathrm{Wi}=4 \& C_{\infty}=0.0$, $\mathrm{Wi}=4 \& C_{\infty}=0.1$, and $\mathrm{Wi}=4 \& C_{\infty}=0.5$ respectively. Dashed lines show the corresponding single-phase stress balance. $\overline{\tau}_N$ is viscous stress , $\overline{\tau}_{R}$ is Reynolds stress, $\overline{\tau}_{P}$ is polymeric stress, $\overline{\tau}_{B}$ is buoyancy stress, $\overline{\tau}_{S}$ is surface stress and $\overline{\tau}_{23}$ is total shear stress. All stress profiles are scaled with the corresponding wall stress $\overline{\tau}_{w}$. (d) Contribution of different stresses to the total friction factor at $\mathrm{Wi}=4$ for various bulk surfactant concentrations.}
\label{StressBV}
\end{figure}

The results are visualized in Fig.~\ref{Qcrit_V} and quantified in Figs.~\ref{SFCV}-\ref{StressBV}. As seen in Fig.~\ref{Qcrit_V}, the clean bubbles move upwards and towards the wall for both the Newtonian and viscoelastic cases, which could be attributed to the combined effects of inertial and elastic lift forces. Note that, the elastic lift force acting on the bubbles is induced by normal stress differences~\cite{mukherjee2013effects,ahmed2020turbulent}. On the other hand, the contaminated ones follow more complicated paths due to additional Marangoni-induced force caused by the non-uniform interfacial surfactant concentration. The inertial lift force and elastic force push the bubbles towards the channel wall, while the Marangoni-induced force opposes them. At $\mathrm{Wi}=4$, initially the elastic and the inertial forces prevail, leading to formation of the wall layers. However, later the Marangoni-induced force becomes sufficiently large resulting in migration of bubbles towards the core region. On the other hand, at $\mathrm{Wi}=8$, the elastic forces become stronger and thus the combined effect of inertial and elastic forces overcome the Marangoni-induced force promoting the formation of wall layers. Note that, in this case, the wall layers are not affected by the surfactants. Figure~\ref{Qcrit_V} shows the vortices for the clean and contaminated viscoelastic turbulent flows, compared with the corresponding Newtonian case. As can be seen, the clean ($C_\infty=0$) Newtonian [Fig.~\ref{Qcrit_V}(a)] and viscoelastic [$\mathrm{Wi}=4$, Fig.~\ref{Qcrit_V}(b)] bubbly flow cases show strong turbulence attenuation and the vortical structures are only concentrated around the bubbles. For the contaminated bubbly flows ($C_\infty=0.1$), the turbulent structures reappear, but are stronger in the Newtonian case [Fig.~\ref{Qcrit_V}(c)] than those in the viscoelastic case [$\mathrm{Wi}=4$, Fig.~\ref{Qcrit_V}(d)]. Increasing concentration to $C_\infty=0.5$ results in less vortical structures especially at $\mathrm{Wi}=8$ where a significant amount of bubbles form wall layer [Fig.~\ref{Qcrit_V}(f)].

\noindent

Although the focus of the present work is on the bubbly flows, we also performed simulations to quantify the effects of viscoelasticity on drag reduction in single-phase flows. For the single-phase viscoelastic base flow, the drag reduction (DR) for $\mathrm{Wi}=4$ and $\mathrm{Wi}=8$ is found to be approximately 23\% and 37\%, respectively, which falls in the LDR regime of polymeric drag reduction~\cite{dubief2004coherent}. The drag reduction for single-phase flow is determined as $DR = (c_{f,0} - c_{f,0,Wi})/c_{f,0}$ where $c_{f,0}$ is the skin friction coefficient for the Newtonian single-phase case and $c_{f,0,Wi}$ is the corresponding viscoelastic case. 

Figure~\ref{SFCV} shows the transient behavior of the skin friction coefficient $(c_f)$ for the clean and the contaminated bubbly flows. The addition of clean bubbles to the viscoelastic turbulent flow has drastic effects on the polymeric drag reduction. As can be seen, $c_f/c_{f,0,Wi}$ increases with the Weissenberg number, whereas $c_f/c_{f,0}$ only slightly changes with $\mathrm{Wi}$. Overall, viscoelasticity results in a significant drag increase in a clean bubbly flow. This is in sharp contrast with the expectations from single-phase flow, where we observe drag reduction by viscoelasticity. Conversely, for the bubbly flow in the presence of surfactant, the drag reduction of viscoelastic turbulent flow is restored, in line with the corresponding single-phase flows. This revival of the drag reduction for the contaminated bubbly flow depends on the net effect of Marangoni, elastic and aerodynamic forces. The loss of drag reduction for clean bubbly flow is due to the formation of bubble wall-layers -- the clean bubbles are pushed toward the channel wall due to combined aerodynamic and elastic lift forces~\cite{ahmed2020effects}. Thus, the addition of polymers into the clean turbulent bubbly flow results in a drag increase due to the elastic induced lift forces which increase with $\mathrm{Wi}$. In the presence of surfactant, the Marangoni forces prevent the formation of bubble clusters by opposing the aerodynamic and elastic induced forces. Thus, for the $\mathrm{Wi}=4 \,\&\, C_\infty=0.1$ case, $c_f$ approaches the corresponding single-phase value. Note that the drag slightly increases with $C_\infty$ which could be attributed to decrease in Marangoni induced force due to more uniform surface concentration~\cite{ahmed2020effects}. For $\mathrm{Wi}=8\,\& \,C_\infty=0.5$ on the other hand, $c_f$ increases again, showing that the Marangoni forces are not strong enough to counter the elastic induced lift force. For viscoelastic turbulent bubbly flow with contamination, the polymer drag reduction hence depends on the intricate interplay of the aerodynamic, Marangoni and elastic forces.

To quantify the combined effects of surfactants and viscoelasticity, time-averaged quantities are plotted in Fig.~\ref{statsV} against the wall units for various single- and multi-phase cases. For single-phase cases in the viscous sublayer ($z^+\leq10$), the profiles follow each other in a linear manner $(v^+ = z^+)$. On the other hand, away from the wall ($z^+\ge30$), the velocity mainly increases with $\mathrm{Wi}$, which is consistent with the decrease in $c_f$. In fact, the mean velocity profiles vary between the von Karman law \cite{pope2001turbulent} for the Newtonian case and the Maximum Drag Reduction (MDR) asymptote ($v^+ =11.7 \ln z^+ -17$)~\cite{virk1970ultimate} for the $\mathrm{Wi}=4$ and $\mathrm{Wi}=8$ cases. As expected, the velocity profile stays parallel to the Newtonian one and move upward with $\mathrm{Wi}$. Moreover, the viscoelastic cases considered in the present work fall in the LDR regime and thus are far from the MDR asysmptote. These findings are consistent with the observations in previous works~\cite{White}. For the multiphase cases, a similar linear variation is observed in the viscous sublayer. In the log-law region, profiles change non-monotonically with $C_\infty$ at $\mathrm{Wi}=4$. For $C_\infty=0$ (clean), the flow presents strong turbulent attenuation. On the other hand, for $C_\infty=0.1$, the profile approaches the corresponding turbulent single-phase case. Increasing further to $C_\infty=0.25$, the profile slightly moves downward and does not change further at $C_\infty=0.5$. Figure~\ref{statsV}(b) shows variation of the average void fraction along the wall normal. As expected for the clean case at $\mathrm{Wi}=4$, there is a clear peak of average void fraction near the wall indicating formation of the wall layer. Moreover, the variation of the peak shows non-monotonic trend with $C_\infty$ at $\mathrm{Wi}=4$. The wall layer is inhibited at $C_\infty=0.1$, whereas it is enhanced at $C_\infty=0.25$ and 0.5. Note that significant peak is restored near the wall at $C_\infty=0.5$ and $\mathrm{Wi}=8$. The formation of bubble wall-layers does not only affect the drag reducing property of polymers, but also causes a significant drag increase for clean bubbly flows. The drag-reducing property of polymers is attributed to the polymer stretching. Figure~\ref{statsV}(c) plots average polymer stretching ($Tr(\bar{B})/L^2$) for $\mathrm{Wi}=4$ and $\mathrm{Wi}=8$ cases. The stretching profile is maximum in the vicinity of the wall and minimum on the centerline. For the single-phase case, the profile moves upwards with $\mathrm{Wi}$. Moreover, the polymeric stretching is highly affected by the formation of bubble clusters. As can be seen, the polymeric stretching becomes minimal for the clean bubbly flow at $\mathrm{Wi}=4$. For the contaminated case at $\mathrm{Wi}=4$, the $Tr(\bar{B})/L^2$ profile coincides with the corresponding single-phase case. Also, it does not change with an increase in $C_\infty$ from 0.1 to 0.5. At $\mathrm{Wi}=8$ and $C_\infty=0.5$, the profile is attenuated with respect to its corresponding single-phase case. This could be attributed to the increased near-wall peak in the void fraction.

To characterize the turbulent flow the liquid velocity fluctuations are shown in Fig.~\ref{RES_V}. For the single-phase case, the peak in the velocity fluctuations increases in the streamwise direction ($v^{\prime}_{rms}$) and decreases in the other directions. Note that the peak of $v^{\prime}_{rms}$ shows slightly non-uniform trend with $\mathrm{Wi}$, i.e., it first increases and then slightly decreases at $\mathrm{Wi}=8$. Indeed, this could be attributed to the fact that the $\mathrm{Wi}=8$ case is approaching the HDR regime. This statistical trend is typical for the LDR and the HDR regimes~\cite{dubief2004coherent,dubief2013mechanism}. On the other hand, the trend for bubbly-flow is more complex. The fluctuations for the clean case are the same as those observed for the Newtonian bubbly flow. For the contaminated case at $\mathrm{Wi}=4$ and $C_\infty=0.1$, there is a little difference between the single-phase case for $u^{\prime}_{rms}$ and $w^{\prime}_{rms}$ profiles, but for $v^{\prime}_{rms}$ profile, the peak is smaller than that in the single-phase case. Further increase in concentration to $C_\infty=0.25$ results in attenuation of fluctuations ($u^{\prime}_{rms}$, $v^{\prime}_{rms}$, $w^{\prime}_{rms}$) away from the wall. Moreover, the reduction of $v^{\prime}_{rms}$ near the wall can be attributed to the increase in the frequency of bubbles leaving/entering in the bubble wall-layer. Indeed, as shown in Fig.~\ref{RES_V}(d), evolution of the number of bubbles in the wall-layer ($N_{wall}$) highly depends on the concentration. Initially, aerodynamic and elastic induced forces dominate, pushing the most of the bubbles to the wall region. The first peak decreases with concentration due to Marangoni forces. As Marangoni forces prevail, $N_{wall}$ starts to decrease. Higher concentration leads to stronger decay in the evolution. The evolution strongly oscillates due to interplay of different forces. In the statistically steady-state, $N_{wall}$ increases with $C_\infty$. Increasing concentration from $C_\infty=0.25$ to $C_\infty=0.5$ at $\mathrm{Wi}=4$ does not show any significant effect in the steady-state. At $\mathrm{Wi}=8$ and $C_\infty=0.5$, after the first peak $N_{wall}$ does not change significantly and oscillates around $N_{wall}/N_{total}	\approx 0.6$. Note that, as we increase the bulk surfactant concentration, the interface becomes more rigid approaching a solid particle~\cite{Tasoglu}. In fact, \citet{esteghamatian2019dilute} have recently found that inclusion of dilute suspension of spherical particles leads to decrease fluctuations in the turbulent viscoelastic flow.   

To examine the drag reduction phenomenon in more details, the shear stress balance across the channel is shown in Fig.~\ref{StressBV}(a-c). Moreover, contribution of different stresses to the friction coefficient is shown on panel (d). For the clean case ($\mathrm{Wi}=4,C_\infty=0$, Fig.~\ref{StressBV}a), the stress balance is completely different from the corresponding single-phase flow (shown by the dashed lines). The surface stress ($\overline{\tau}_{S}$) and buoyancy stress ($\overline{\tau}_{B}$) have significant contribution to the total shear stress ($\overline{\tau}_{23}$). Due to the significant reduction in the flow rate, the viscous stress ($\overline{\tau}_N$) near the wall reduces. The polymeric stress ($\overline{\tau}_{P}$) almost vanishes for the clean bubbly flow. Similar to the Newtonian case, the Reynolds stress also significantly diminishes due to presence of wall layers. Conversely, for the $\mathrm{Wi}=4$ and $C_\infty=0.1$ case, the stress balance becomes similar to the corresponding single-phase flow, as the major contribution to the total shear stress ($\overline{\tau}_{23}$) comes from the Reynolds stress, the viscous stress and polymeric stress. In comparison to the clean case at $\mathrm{Wi}=4$, the surface and the buoyancy stresses significantly reduce, while the polymeric stress increases. Also, the Reynolds stress is slightly lower for the contaminated bubbly flow than that in the corresponding single-phase flow, and further decreases for higher surfactant concentrations ($\mathrm{Wi}=4,C_\infty=0.5$). As more bubbles migrate towards the wall at $C_\infty=0.5$, the surface and the buoyancy stresses prevail and the Reynolds stresses attenuate, especially at $\mathrm{Wi}=8$ and $C_\infty=0.5$ (not shown here).

\section{Conclusions}\label{sec:conclusions}
Extensive interface-resolved direct numerical simulations have been performed to examine the effects of clean and contaminated bubbles on Newtonian and viscoelastic turbulent channel flows. It is found that the formation of bubble-rich wall layers plays a major role in turbulent bubbly flows and its prevention is of crucial importance in realizing the polymer drag reduction. For the clean case, the earlier findings are verified: bubbles move towards the wall due to the hydrodynamic lift force, and form a dense wall layer, which dramatically increases the drag and completely suppresses the polymer drag reduction. Prevention of bubble clusters near the wall by surfactant has been also examined by using the sorption properties of Triton X-100 and 1-Pentanol. It is found that the sorption kinetics highly affects the dynamics of turbulent bubbly flow. For Triton X-100, a minute amount of surfactant (e.g., $C_\infty=0.1$)  is found to be sufficient to prevent the formation of bubble clusters while, for 1-Pentanol, even $C_\infty=1.0$ is not high enough. The flow rate for both $\mathrm{Wi}=4$ and $\mathrm{Wi}=8$ is similar to that of the Newtonian ($\mathrm{Wi}=0$) clean bubbly flow. We also showed that the addition of small amount of surfactant (contaminated bubbles) can revive the polymer drag reduction effect for turbulent bubbly flows. It is found that viscoelasticity promotes formation of the bubble-wall layer. Thus the benefit of drag reduction by polymers in single phase flow is completely lost in the bubbly flows unless a strong enough surfactant is added to the system. Drag reduction for contaminated bubbly flows depends on the intricate interplay of the hydrodynamic, elastic and Marangoni-induced forces. For $\mathrm{Wi}=4$ and $C_\infty=0.5$, almost the same drag reduction was achieved in bubbly flow as in the respective single-phase flow, but for $\mathrm{Wi}=8$ and $C_\infty=0.5$, the drag reduction did not approach its single-phase value. The reason for the difference can be found in the elastic-induced lift force, which increases with $\mathrm{Wi}$, and promotes the formation of bubble-wall layers. Marangoni-induced force, on the other hand, pushes bubbles to the center of the channel and counteracts the elastic-induced lift force. At $\mathrm{Wi}=4$, Marangoni-induced force is sufficient to balance the elastic-induced lift force, but at $\mathrm{Wi}=8$, the bubbly wall layers started to reappear.

\section*{Acknowledgments}
We acknowledge financial support by the Swedish Research Council through grant No. VR2017-4809. This project also has received funding from the European Research Council
(ERC) under the European Union's Horizon 2020 research and innovation programme (StG MUCUS, No. 852529). This effort receives funding by the Aalto Science Foundation (ASCI) for funding the PENGUIN project. MM acknowledges financial support from TUBITAK grant No.115M688. We acknowledge PRACE for awarding us access to JUWELS at GCS@FZJ, Germany. Additional computational time was provided by SNIC (Swedish National Infrastructure for Computing). PC acknowledges funding from the University of Iceland Recruitment Fund grant no. 1515-151341, TURBBLY.
\pagebreak
\bibliography{IzbassarovetalPRF}
\end{document}